\documentclass[proof]{WileyASNA-v1}

\articletype{Article Type}%

\received{?? ???? 2024}
\revised{?? ???? 2024}
\accepted{?? ???? 2024}

\newcommand\arcmin{\hbox{$^\prime$}}

\newcommand\scalemath[2]{\scalebox{#1}{\mbox{\ensuremath{\displaystyle #2}}}}

\begin{document}

\title{Photometric and kinematic studies of open clusters Ruprecht 1 and Ruprecht 171}

\author[1]{H. \c{C}akmak}
\author[2]{T. Yontan}
\author[2]{S. Bilir}
\author[3,4]{T. S. Banks}
\author[5]{R. Michel}
\author[6,7]{E. Soydugan}
\author[8]{S. Ko\c{c}}
\author[8]{H. Er\c{c}ay}

\authormark{\c{C}akmak \textit{et al.}}

\address[1]{\orgdiv{Faculty of Science, Department of Computer Sciences}, \orgname{Istanbul University}, \orgaddress{\state{Istanbul}, \country{Türkiye}}}

\address[2]{\orgdiv{Faculty of Science, Department of Astronomy and Space Sciences}, \orgname{Istanbul University}, \orgaddress{\state{Istanbul}, \country{Türkiye}}}

\address[3]{\orgdiv{Nielsen, 675 6th Ave., NYC}, \orgaddress{\state{NY}, \country{USA}}}

\address[4]{\orgdiv{Harper College, 1200 W Algonquin Rd, Palatine}, \orgaddress{\state{Illinois}, \country{USA}}}

\address[5]{\orgdiv{Universidad Nacional Autonoma de Mexico}, \orgname{Observatorio Astronomico Naciona}, \orgaddress{\state{Ensenada}, \country{Mexico}}}

\address[6]{\orgdiv{Faculty of  Sciences, Department of Physics}, \orgname{\c{C}anakkale Onsekiz Mart University}, \orgaddress{\state{\c{C}anakkale}, \country{Türkiye}}}

\address[7]{\orgdiv{Astrophysics Research Center and Ulupınar Observatory}, \orgname{\c{C}anakkale Onsekiz Mart University}, \orgaddress{\state{\c{C}anakkale}, \country{Türkiye}}}

\address[8]{\orgdiv{Institute of Graduate Studies in Science}, \orgname{Istanbul University}, \orgaddress{\state{Istanbul}, \country{Türkiye}}}

\corres{*Hikmet \c{C}akmak, Faculty of Science, Department of Computer Sciences, Istanbul, Türkiye. \email{hcakmak@istanbul.edu.tr}}


\abstract{This study outlines a detailed investigation using CCD {\it UBV} and {\it Gaia} DR3 data sets of the two open clusters Ruprecht 1 (Rup-1) and Ruprecht 171 (Rup-171). Fundamental astrophysical parameters such as color excesses, photometric metallicities, ages, and isochrone distances were based on {\it UBV}-data analyses, whereas membership probability calculations, structural and astrophysical parameters, as well as the kinematic analyses were based on {\it Gaia} DR3-data. We identified 74 and 596 stars as the most probable cluster members with membership probabilities over 50\% for Rup-1 and Rup-171, respectively. The color excesses $E(B-V)$ were obtained as $0.166\pm0.022$ and $0.301\pm0.027$ mag for Rup-1 and Rup-171, respectively. Photometric metallicity analyses were performed by considering F-G type main-sequence member stars and found to be [Fe/H]=$-0.09\pm 0.16$ and [Fe/H]=$-0.20\pm 0.20$ dex for Rup-1 and Rup-171, respectively. Ages and distances were based on both {\it UBV} and {\it Gaia}-data analyses; according to isochrone-fitting these values were estimated to be $t=580\pm60$ Myr, $d=1469\pm57$ pc for Rup-1 and $t=2700\pm200$ Myr, $d=1509\pm69$ pc for Rup-171. The present-day mass function slope of Rup-1 was estimated as $1.26\pm0.32$ and Rup-171 as $1.53\pm1.49$. Galactic orbit integration analyses showed that both of the clusters might be formed outside the solar circle.}


\jnlcitation{\cname{%
\author{H. \c{C}akmak},
\author{T. Yontan},
\author{S. Bilir},
\author{T. Banks},
\author{M. Ra\'{u}l},
\author{E. Soydugan},
\author{S. Ko\c{c}}, and
\author{H. Er\c{c}ay}%
} (\cyear{2024}), 
\ctitle{Photometric and kinematic studies of open clusters Ruprecht 1 and Ruprecht 171}, \cjournal{Astronomische Nachrichten}, \cvol{2024;00:1--19}.}


\maketitle
\section{Introduction} \label{sec:intro}
The study of open star clusters (OCs) in our Galaxy can offer valuable insights. OCs are loose groupings of stars, bound together by their (weak) self-gravitational force. OCs contain stars of similar age and composition making them, for example, excellent laboratories for studying stellar evolution. Metal abundance, distance, kinematics, and age can be estimated leading to Galactic OCs acting as tracers into the structure, formation, and evolution (in both chemistry and structure) of the Galactic disk \citep{Friel95}. The current paper is part of a wider project using a common methodology across detailed studies of OCs \citep[see][and references therein]{Yontan15, Yontan19, Ak16, Yontan23a}, making detailed and careful analyses of otherwise neglected OCs and building towards a meta-analysis. 

The high-precision astrometric, photometric, and spectroscopic data of the {\it Gaia} space mission provides a foundation for high-quality astrophysics research \citep{Gaia16}. The astrometric data from this mission makes identification of the cluster members easier \citep[e.g.,][]{vanLeeuwen22, Sethi23}. Many researchers have successfully performed membership analyses from the proper-motions and trigonometric parallaxes of {\it Gaia} \citep[e.g.,][]{Cantat-Gaudin18, Bostanci18, Bisht21, Yontan21, Yontan22}. Such clearly distinguished groups made up of cluster members supply cleaner color-magnitude and color-color diagrams, as well as allow more accurate calculations of the fundamental astrophysical parameters for the clusters under study. 

The mass function of OCs highlights the diversity and dynamics of stellar populations. As a group of stars formed from the same molecular cloud and typically represent a wide range of stellar masses, OCs are useful tools to study present-day and initial mass functions. Various authors have investigated these functions for OCs, exploring whether the initial-mass function is universal for all OCs or if it is affected by star-forming processes \citep[e.g.,][]{Kroupa02, Dib17, Joshi20}. The study of OCs gives insight into the topics of dynamical evolution, mass segregation for OCs, and thus the distribution of different stellar masses in the clusters \citep[e.g.,[]{Bisht19, Bisht21}. 

\subsection{Ruprecht~1}
\begin{figure*}[h!]
	\centering
	\includegraphics[width=0.8\linewidth]{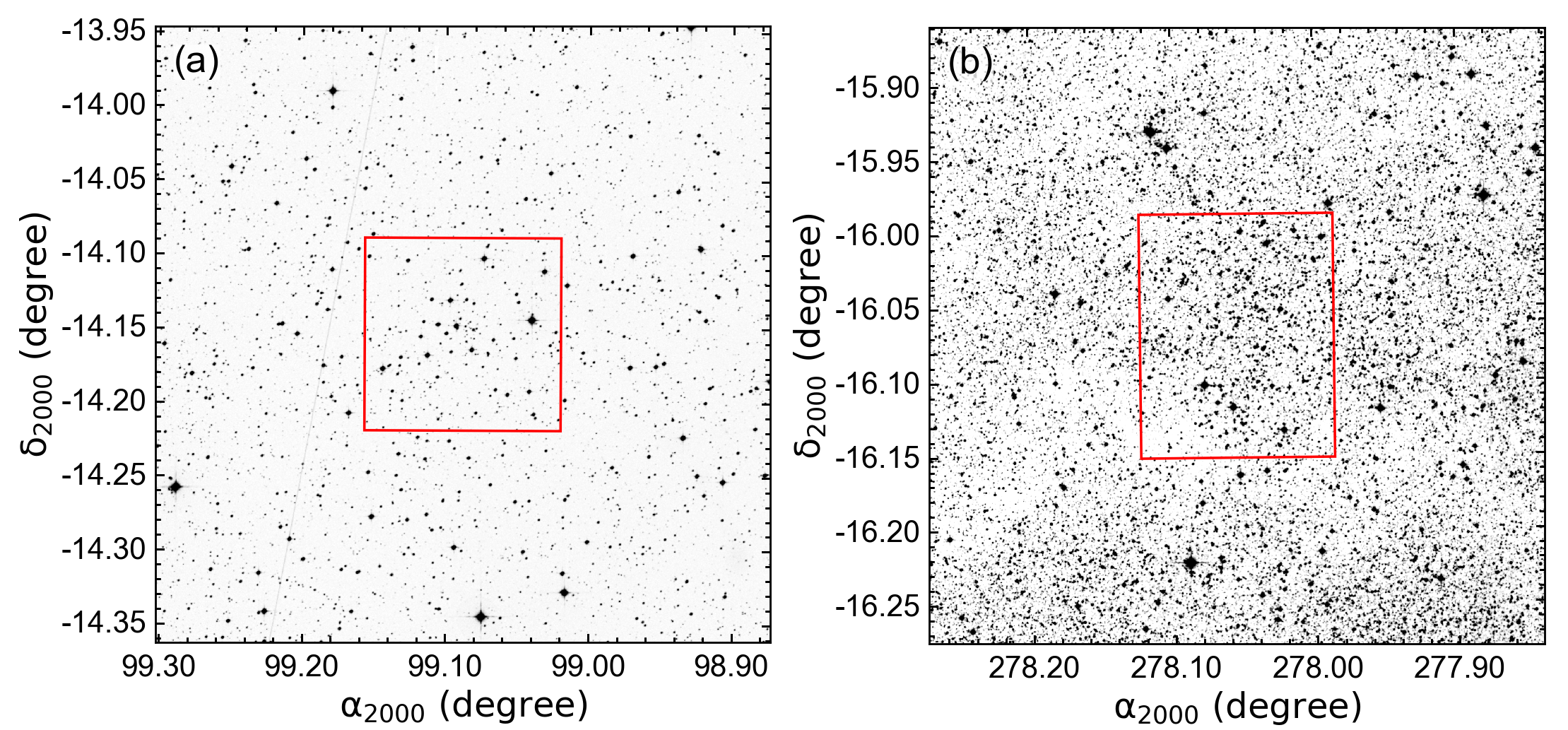}
	\caption{Identification charts for Rup-1 (a) and Rup-171 (b). Red boxes represent the {\it UBVRI} field of view of $7^{'}\!\!.6\times 7^{'}\!\!.6$ for Rup-1 (a) and $7^{'}\!\!.6\times 9^{'}\!\!.2$ for Rup-171 (b). The field of view of the charts are $25^{'}\times 25^{'}$. North is up and East is left. The charts are taken from the STScI Digitized Sky Survey (https://archive.stsci.edu/cgi-bin/dss\_form)} 
	\label{fig:ID_charts}
\end{figure*}

\citet{Ruprecht66} presented the cluster Ruprecht~1 ($\alpha =$ 06:36:20.2, $\delta =$ $-$14:09:25, J2000), assigning it a \citet{Trumpler_1930} classification of `III 1 p', indicating a poorly populated detached cluster with no concentration, composed of less than 50 (then) observed stars having nearly the same apparent brightness. The identification chart of this cluster is shown in Fig.~\ref{fig:ID_charts}-a. \citet{Kharchenko05} included the cluster in their catalog of astrophysical data for 520 Galactic OCs. The cluster was estimated to have an angular radius of $15 \arcmin$, $E(B-V)=0.15$ mag, a distance of 1100 pc, and an age of 575 Myr. These values contrast with the results of \citet{Piatti08}, who made CCD observations of the cluster using the Washington $C$ and the Kron-Cousins $R_{\rm KC}$ (in place of Washington $T_1$) bands. \citet{Piatti08} estimated the reddening $E(B-V)$ as $0.25 \pm 0.05$ mag, the apparent radius as $5^{'}\!\!.3 \pm 0^{'}\!\!.4$ (and hence the physical radius as $\sim 2.6 \pm 0.2$ pc), and provided upper and lower estimates for the distance and age by assuming first an upper limit of $z=0.02$ and a lower one of 0.008. The cluster distance was therefore estimated as between $1.9 \pm 0.4$ and $1.5 \pm 0.3$ kpc, and the cluster age as between $200 \pm 47$ and $251 \pm 58$ Myr. \citet{Piskunov_2007} fitted \citet{King62} models to 236 OCs listed in the catalog of \citet{Kharchenko05}, and so estimated core and tidal radii as well as the tidal masses of the studied clusters. These authors derived Rup-1's  core radius as 2.1 pc, the tidal radius as 4.6 pc, and the log cluster mass (in solar units) as 1.462. Subsequently \citet{Piskunov_2008} built off \citet{Piskunov_2007}, revising the tidal radius to 7.6 pc and the logarithmic cluster mass to 2.554 solar units. Later papers, such as \citet{Dias14, Oralhan15, Sampedro17, Loktin17, Cantat-Gaudin18, Bossini19, Cantat-Gaudin20, Dias21}, included Ruprecht~1 in large scale analyses of many clusters, with \citet{Piatti08} being the last in-depth study of the cluster. Table~\ref{tab:literature} presents the key results of these studies and shows that there is still a spread in the estimates (with values often being copied over from earlier studies). Hence, as aimed for in this study, detailed analyses should be performed to clarify parameters for the cluster.

\begin{table*}[h!]
	\setlength{\tabcolsep}{10pt}
	\renewcommand{\arraystretch}{1.0}
	\fontsize{8pt}{10pt}\selectfont
	\centering
	\caption{Summary of results from the literature for the Rup-1 and Rup-171 OCs. Columns are color excess ($E(B-V$)), distance ($d$), iron abundance ([Fe/H]), method for determination of metal abundance (Method), age ($t$), proper-motion components ($\langle\mu_{\alpha}\cos\delta\rangle$, $\langle\mu_{\delta}\rangle$), and radial velocity ($V_{\rm R}$). `Ref' indicates the source of the data, according to the list of papers below the table.}
	\begin{tabular}{ccccccccl}
		\toprule
		\multicolumn{9}{c}{Rup-1}\\
		\toprule
		$E(B-V)$ &  $d$ & [Fe/H] & Method{$^*$} &$t$ &  $\langle\mu_{\alpha}\cos\delta\rangle$ &  $\langle\mu_{\delta}\rangle$ & $V_{\rm R}$ & Ref \\
		(mag) &  (pc)  & (dex) & & (Myr) & (mas yr$^{-1}$) & (mas yr$^{-1}$) & (km s$^{-1})$ &      \\
		\bottomrule
		\toprule
		0.15            & 1100          & ---              & --- & 575        & +2.20$\pm$0.48     & 0.10$\pm$0.34      & ---             & (01) \\
		0.25$\pm$0.05   & 1900$\pm$400  & 0.00             & Pho         & 200$\pm$47 & ---                & ---                & ---             & (02) \\
		0.25$\pm$0.05   &  1500$\pm$300 & $-$0.40          & Pho         & 251$\pm$8  & ---                & ---                & ---             & (02) \\
		0.15            &  1100         & ---              & --- & 575        & ---                & ---                & ---             & (03) \\
		0.146           &  1204         & ---              & --- & 615        & 0.48               & 0.44               & ---             & (04) \\
		---             &  ---          & ---              & --- & ---        & 0.49$\pm$0.10      & $-$0.99$\pm$0.83   & ---             & (05) \\
		0.17$\pm$0.06   &  1480$\pm$30  & $-$0.25$\pm$0.18 & Pho         & 480$\pm$40 & ---                & ---                & ---             & (06) \\
		0.26            &  1720         & ---              & --- & 500        & 0.11$\pm$4.08      & $-$1.12$\pm$4.16   & ---             & (07) \\
		0.088           &  2409         & ---              & --- & 630        & 0.064$\pm$0.082    & 0.884$\pm$0.104    & ---             & (08) \\
		---             &  ---          & ---              & --- & ---        & $-$0.94$\pm$0.16   & 0.20$\pm$1.25      & ---             & (09) \\
		---             &  1514$\pm$235 & ---              & --- & ---        & $-$0.316$\pm$0.012 & 0.900$\pm$0.010    & ---             & (10) \\
		---             &  1514$\pm$235 & ---              & --- & ---        & $-$0.316$\pm$0.012 & 0.900$\pm$0.010    & 12.52$\pm$3.70  & (11) \\
		---             &  1563$\pm$105 & ---              & --- & 576$\pm$35 & $-$0.305$\pm$0.171 & $-$0.867$\pm$0.227 & ---             & (12) \\
		0.225$\pm$0.007 &  1437$\pm$16  & $-$0.25          &Pho          & 320$\pm$60 & ---                & ---                & ---             & (13) \\
		0.197           &  1533         & ---              & ---              & 290        & $-$0.316$\pm$0.098 & $-$0.900$\pm$0.103 &  ---            & (14) \\
		0.240$\pm$0.024 & 1429$\pm$68   &$-$0.077$\pm$0.094& Pho         & 430$\pm$80 & $-$0.322$\pm$0.131 & $-$0.899$\pm$0.118 & 11.46$\pm$0.56  & (15) \\
		---             &  1480         & ---              & --- & 310        & ---                & ---                & 12.52$\pm$3.70  & (16) \\
		0.083$\pm$0.031 &  1462$\pm$5   & ---              & --- & 487$\pm$209& $-$0.293$\pm$0.053 & $-$0.901$\pm$0.062 & ~9.39$\pm$1.94  & (17) \\
		0.166$\pm$0.022 &  1469$\pm$57  & -0.09$\pm$0.16   & Pho         & 580$\pm$60& $-$0.287$\pm$0.003  &$-$0.903$\pm$0.003 & 10.37$\pm$2.22  & (18) \\
		\bottomrule
		\toprule
		\multicolumn{9}{c}{Rup-171}\\
		\toprule
		0.12            &  1140$\pm$50  & ---              & --- & 3200        & ---               & ---                & ---             & (19) \\
		0.25            &  1159         & ---              & --- & 3550        & ---               & ---                & ---             & (04) \\
		---             &  ---          & ---              & --- & ---         & 1.58$\pm$4.01     & 0.32$\pm$1.52      & ---             & (05) \\
		0.12            &  1140         & ---              & --- & 3160        & 1.11$\pm$6.42     & 0.17$\pm$5.91      & ---             & (07) \\
		---             &  1540$\pm$243 & ---              & --- & ---         & 7.677$\pm$0.008   & 1.091$\pm$0.008    & ---             & (10) \\
		---             &  1514$\pm$235 & ---              & --- & ---         & 7.677$\pm$0.008   & 1.091$\pm$0.008    & 5.61$\pm$0.22   & (11) \\
		---             &  1577$\pm$132 & ---              & --- & 4$\pm$0.2   & 7.680$\pm$0.121   & 1.096$\pm$0.207    & ---             & (12) \\
		0.219           &  1522$\pm$162 & 0.06$\pm$0.03    & Spe          & 2750              & 7.677$\pm$0.008   & 1.091$\pm$0.008    & ---             & (20) \\
		0.219           &  1522$\pm$162 & ---              & --- & 2750        & 7.677$\pm$0.088   & 1.091$\pm$0.008    & ---             & (14) \\
		0.298$\pm$0.019 &  1512$\pm$15  & 0.048$\pm$0.057  & Pho          & 2960$\pm$145      & 7.678$\pm$0.226   & 1.091$\pm$0.210    & 5.488$\pm$0.57  & (15) \\
		---             &  1476         & $-0.041\pm0.014$ & Spe          & 2820$\pm$50       & ---               & ---                & ---             & (21) \\
		---             &  1476         & ---              & --- & 2820        & ---               & ---                & 5.70$\pm$0.17   & (16) \\
		0.262$\pm$0.059 &  1482$\pm$3   & ---              & --- & 1472$\pm$603& 7.717$\pm$0.130   & 1.083$\pm$0.013    & 5.56$\pm$1.16   & (17) \\
		0.301$\pm$0.027& 1509$\pm$69   & -0.20$\pm$0.20    & Pho & 2700$\pm$200& 7.720$\pm$0.002   & 1.082$\pm$0.002    & 5.32$\pm$0.23   & (18) \\
		\bottomrule
		\toprule
	\end{tabular}%
	\\
	\raggedright
	\vspace{5pt}
	(01)~\citet{Kharchenko05}, (02)~\citet{Piatti08}, (03)~\citet{Kharchenko09}, (04)~\citet{Kharchenko13}, (05)~\citet{Dias14}, (06)~\citet{Oralhan15}, (07)~\citet{Sampedro17}, (08)~\citet{Loktin17}, (09)~\citet{Dias18}, (10)~\citet{Cantat-Gaudin18}, (11)~\citet{Soubiran18}, (12)~\citet{Liu19}, (13)~\citet{Bossini19}, (14)~\citet{Cantat-Gaudin20}, (15)~\citet{Dias21}, (16)~\citet{Tarricq21}, (17)~\citet{Hunt23}, (18)~This study, (19)~\citet{Tadross08}, (20)~\citet{Casali20}, (21)~\citet{Casamiquela21} 
\\
($^{*}$)Pho: Photometric, Spe: Spectroscopic
	\label{tab:literature}%
\end{table*}

\subsection{Ruprecht 171}
\citet{Ruprecht66} classified Ruprecht 171 ($\alpha = $18:32:02.9, $\delta =$ $-$16:03:43, J2000) as `II 1 m', or a detached cluster with little noticeable concentration, with a medium number of stars (in the range 50 to 100 inclusive) of the same apparent brightness.  The identification chart of this cluster is shown in Fig.~\ref{fig:ID_charts}\negmedspace\negmedspace-b. Many of the same catalog studies noted above for Ruprecht~1 included this system, as listed in Table~\ref{tab:literature}. Again, we see the repetition of parameter estimates from catalog to catalog together with a scatter in the independent estimates. A key additional paper is that of \citet{Casali20}, who examined {\it Gaia} DR2 data accompanied by high-resolution optical spectra of seven red giant branch and red clump stars assessed to have a high probability for cluster membership. The estimates for distance and [Fe/H] were in reasonable agreement with those of \citet{Casamiquela21} due to the large estimated uncertainties, as are those of \citet{Dias21}. However, reddening varies substantially across the literature estimates, with even recent outliers for age \citep{Liu19}. As noted for Ruprecht 1, this cluster seems to be in need of additional study.

The current paper explores and characterizes two clusters, Ruprecht-1 (hereafter Rup-1) and Ruprecht-171 (hereafter Rup-171). In this study, CCD {\it UBV} photometric and {\it Gaia} DR3 astrometric, photometric, and spectroscopic data were used together for the first time to investigate Rup-1 and Rup-171. During the analyzes, we considered two separate catalogs for each cluster: a {\it UBV} catalog that contained magnitude and color measurements (see later for details), and the {\it Gaia} catalog gathered from the {\it Gaia} DR3 database which included the stars located in 25 arcmin areas from each cluster center and was comprised of these stars' astrometric, photometric, and spectroscopic measurements. The membership probabilities of stars were calculated from {\it Gaia} catalog. Then we cross-matched the two catalogues, allowing the membership probabilities of the same stars in the {\it UBV} catalog to be determined. The {\it UBV} based catalog was used to obtain fundamental astrophysics parameters such as $E(B-V)$ and $E(U-B)$ color excesses, photometric metallicities [Fe/H], ages, and isochrone distances of the two OCs. The {\it Gaia}-based catalog was used in the estimation of structural and astrophysical parameters, as well as to investigate their astrometric, dynamic, and kinematic properties. In this study we investigated the two OCs in detail performing individual methods that are described in later sections. Hence, we aimed to determine homogeneous results and eliminate the uncertainties in the cluster parameters given in literature.  

\begin{sidewaystable*}[ph!]
	\renewcommand{\arraystretch}{1.4}
	\fontsize{8pt}{10pt}\selectfont
	\captionsetup{labelfont=bf, font=normal}
	\centering
	\caption{The photometric and astrometric catalogs for Rup-1 and Rup-171.}
	\label{tab:input_parameters}
	\begin{tabular}{cccccccccccc}
		\hline
		\multicolumn{11}{c}{Ruprecht 1}\\
		\hline
		ID	 & R.A.           &	Decl.	       &      $V$	    &	$U-B$ & $B-V$	&	$G$	 & $G_{\rm BP}-G_{\rm RP}$	& $\mu_{\alpha}\cos\delta$ & $\mu_{\delta}$ & $\varpi$ & $P$ \\
		& (hh:mm:ss.ss)  &(dd:mm:ss.ss)	  &      (mag)	   &    (mag)& (mag)   &  (mag) & 	(mag)	               & (mas yr$^{-1}$)          & (mas yr$^{-1}$)& (mas)	  &     \\
		\hline
		001 & 06:36:04.84 & -14:06:05.91 & 16.466(0.007) & 0.189(0.013)  & 0.697(0.009) & 16.285(0.003) & 0.959(0.010) & -0.322(0.043) & 3.047(0.049)  & 0.268(0.051) & 0.41 \\
		002 & 06:36:04.90 & -14:08:11.89 & 19.292(0.031) & 0.186(0.078)  & 0.812(0.046) & 19.104(0.004) & 1.041(0.042) & -0.081(0.216) & 1.296(0.253)  &-0.089(0.253) & 0.24 \\
		003 & 06:36:05.21 & -14:13:07.18 & 18.945(0.023) & 0.798(0.103)  & 1.056(0.035) & 18.647(0.003) & 1.351(0.049) & -0.652(0.164) & 0.630(0.171)  & 0.334(0.161) & 0.09 \\
		004 & 06:36:05.22 & -14:12:01.84 & 20.212(0.060) &  -----        & 1.192(0.094) & 19.773(0.005) & 1.674(0.099) & -1.516(0.338) & 0.492(0.361)  & 0.116(0.346) & 0.05 \\
		005 & 06:36:05.30 & -14:09:53.34 & 15.459(0.006) & 0.151(0.011)  & 0.732(0.010) & 15.260(0.003) & 0.977(0.005) & -0.435(0.026) &-0.618(0.029)  & 0.344(0.029) & 0.45 \\
		... & ... & ...	& ... & ... & ...& ... & ... & ... & ... & ...  & ...\\
		182 & 06:36:37.25 & -14:10:51.13 & 19.991(0.044) & -----         & 0.804(0.062) & 19.684(0.005) & 1.107(0.092) &  0.431(0.362) & 0.723(0.396)  & 0.054(0.419) & 0.03 \\
		183 & 06:36:37.28 & -14:11:08.88 & 18.000(0.012) & 1.168(0.073)  & 1.121(0.022) & 17.569(0.003) & 1.433(0.022) & -2.095(0.086) &-1.024(0.095)  & 0.543(0.108) & 0.03 \\
		184 & 06:36:37.51 & -14:07:44.27 & 20.105(0.048) & -----         & 0.594(0.070) & 19.815(0.005) & 0.918(0.099) & -0.088(0.366) &-0.049(0.390)  & 1.518(0.437) & 0.38 \\
		185 & 06:36:37.54 & -14:08:57.51 & 18.656(0.019) & 1.269(0.150)  & 1.213(0.029) & 18.136(0.003) & 1.630(0.032) &  2.184(0.138) &-4.449(0.172)  & 0.677(0.155) & 0.00 \\
		186 & 06:36:37.56 & -14:09:12.87 & 19.674(0.048) & 0.454(0.127)  & 0.790(0.058) & 19.316(0.004) & 1.290(0.055) &  1.332(0.256) &-0.612(0.275)  & 0.057(0.301) & 0.23 \\
		\hline
            \\[5ex]
            \hline
             \multicolumn{11}{c}{Ruprecht 171}\\
		\hline
		ID	 & R.A.           &	Decl.	       &      $V$	    &	$U-B$ & $B-V$	&	$G$	 & $G_{\rm BP}-G_{\rm RP}$	& $\mu_{\alpha}\cos\delta$ & $\mu_{\delta}$ & $\varpi$ & $P$ \\
		& (hh:mm:ss.ss)  &(dd:mm:ss.ss)	  &      (mag)	   &    (mag)& (mag)   &  (mag) & 	(mag)	               & (mas yr$^{-1}$)          & (mas yr$^{-1}$)& (mas)	  &     \\
		\hline
		001 & 18:31:57.23 & -16:08:43.07 & 18.455(0.036) & 1.621(0.231)  & 1.329(0.049) & 18.076(0.003) & 1.673(0.030) & 7.707(0.154) &  1.291(0.128) & 0.690(0.141) & 1.00 \\
		002 & 18:31:57.44 & -16:07:01.19 & 18.573(0.054) & -----         & 1.360(0.077) & 18.431(0.004) & 1.779(0.041) & 0.854(0.229) & -1.378(0.186) & 0.067(0.200) & 0.07 \\
		003 & 18:31:57.56 & -16:06:47.73 & 14.999(0.008) & 0.157(0.016)  & 0.750(0.016) & 14.793(0.003) & 1.009(0.005) & 7.696(0.031) &  0.941(0.027) & 0.633(0.027) & 1.00 \\
		004 & 18:31:58.37 & -16:08:59.33 & 15.393(0.010) & 0.182(0.010)  & 0.662(0.012) & 15.083(0.003) & 1.028(0.005) & 7.755(0.032) &  0.998(0.027) & 0.649(0.030) & 1.00 \\
		005 & 18:31:58.38 & -16:08:23.09 & 14.741(0.008) & 0.228(0.011)  & 0.735(0.012) & 14.482(0.003) & 1.037(0.005) & 0.558(0.027) & -1.509(0.022) & 0.685(0.025) & 0.03 \\
		... & ... & ...	& ... & ... & ... & ... & ... & ... & ... & ...  & ...\\
		366 & 18:32:29.26 & -16:03:26.12 & 17.544(0.016) & 0.311(0.047)  & 1.065(0.035) & 17.155(0.003) & 1.293(0.013) & 1.742(0.087) &  0.401(0.072) & 0.408(0.082) & 0.13 \\
		367 & 18:32:29.41 & -16:04:38.30 & 18.648(0.040) & 0.877(0.169)  & 1.445(0.066) & 18.134(0.004) & 1.848(0.026) &-4.376(0.275) & -6.076(0.251) & 0.657(0.262) & 0.00 \\
		368 & 18:32:29.44 & -16:04:16.85 & 16.773(0.039) & -----         & 1.004(0.047) & 16.370(0.003) & 1.486(0.011) &-0.466(0.074) & -2.774(0.060) & 0.247(0.074) & 0.07 \\
		369 & 18:32:29.46 & -16:08:55.69 & 17.784(0.015) & 0.896(0.064)  & 1.176(0.023) & 17.331(0.003) & 1.474(0.017) &-3.038(0.098) & -1.982(0.081) & 0.675(0.100) & 0.00 \\
		370 & 18:32:29.47 & -16:08:23.28 & 18.406(0.025) & -----         & 1.902(0.051) & 17.254(0.003) & 2.458(0.028) & 1.022(0.094) & -2.644(0.079) & 0.481(0.098) & 0.17 \\
		\hline
	\end{tabular}
	\label{tab:all_cat}%
\end{sidewaystable*}

\begin{figure*}
	\centering
	\includegraphics[width=0.9\linewidth]{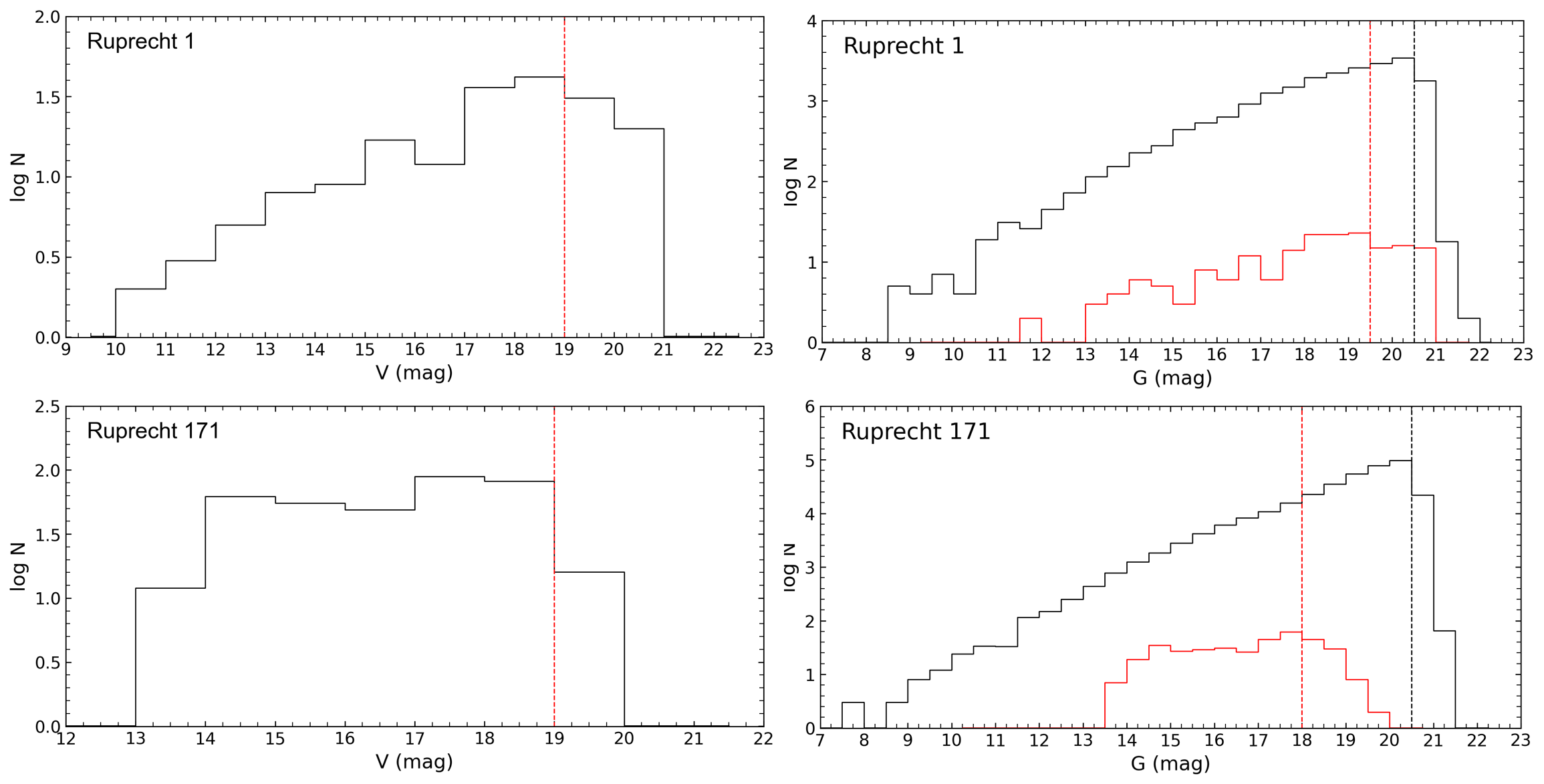}\\
	\caption{Interval $V$-band (left panels) and $G$-band (right panels) magnitude histograms of Rup-1 and Rup-171. The black and red histograms in the right panels are based on the {\it Gaia} DR3 data of stars gathered for the $25'\times25'$ cluster regions (black) and those observed in the {\it UBV} bands (red). The black dashed lines represent the faint $G$ magnitude limit for observed stars, whereas the red dashed lines show the faint apparent magnitude limits in $V$ bands considered in the study.} 
	\label{fig:histograms}
\end{figure*}
\section{Observations and data reductions}
The observations of these two clusters were carried out at the San Pedro Martir Observatory,\footnote{\url{https://www.astrossp.unam.mx/en/users/telescopes/0-84m-telescope}} as part of an ongoing {\it UBVRI} photometric survey of Galactic stellar clusters started on September 2009.  Up to date 1,496 observations of  \href{http://www.astrosen.unam.mx/~rmm/SPMO_UBVRI_Survey/Clusters_Open.html}{1,385 open clusters} and 149 observations of \href{http://www.astrosen.unam.mx/~rmm/SPMO_UBVRI_Survey/Clusters_Globular.html}{87 globular clusters} have been carried out. The publication of the details of this survey is in preparation by Ra\'ul Michel. The 84 cm ($f/15$) Ritchey-Chretien telescope was employed in combination with the Mexman filter wheel.

Rup-1 was observed on 2016-11-07 with the Marconi 3 detector (a $2048 \times 2048$ 13.5-$\mu$m square-pixels e2v CCD42-40 with a gain of $1.71 \:\mathrm{e^-}$ ADU$^{-1}$ and a readout noise of $4.9 \: \mathrm{e^-}$, giving a field of view of about $7.6 \times 7.6$ arcmin$^2$). Short and long exposures were taken to properly measure both the bright and faint stars of the fields. Exposure times for {\it I} and {\it R} were 2, 20, 200s in duration; 4, 40, 400s for {\it V}; 6, 60, 600s for {\it B}; and 10, 100, 1000s for {\it U}.

Rup-171 was observed on 2013-06-09 with the ESOPO CCD detector (a $2048 \times 4612$ 13.5-$\mu$m square-pixels e2v CCD42-90 with a gain of $1.83 \: \mathrm{e^-}$ ADU$^{-1}$ and a readout noise of $4.7 \: \mathrm{e^-}$ at the $2 \times 2$ binning employed, providing an unvignetted field of view of about $7.6 \times 9.2$ arcmin$^2$). Three different exposure times per filter were used without stacked images at all. Exposure times were 10, 50, 200s for both {\it I} and {\it R}; 10, 20, 200 for {\it V}; 10, 20, 300s for {\it B}; and 30, 60, 600s for {\it U}.

The observations were carried out during very photometric conditions. Landolt's standard stars \citep{Landolt09} were also observed, at the meridian and at about two airmasses, to properly determine the atmospheric extinction coefficients. Flat fields were taken at the beginning and the end of each night and bias images were obtained between cluster observations. Data reduction with point spread function (PSF) photometry was carried out by Ra\'ul Michel with the IRAF/DAOPHOT packages \citep{Stetson87} and employing the transformation equations recommended, in their Appendix B, by \citet{Stetson19}.

\section{Data Analysis}
\subsection{UBV Photometric Data}
Data reduction and analyses resulted in {\it UBV} photometric catalogs of 186 and 370 stars for Rup-1 and Rup-171, respectively (Table \ref{tab:input_parameters}). The coordinate solution for the targets was performed using the astrometry packages of IRAF. These catalogs contain equatorial coordinates, $V$-band magnitudes and $U-B$, $B-V$ color indices, and relevant photometric errors of each detected star. $V$-band magnitudes of the stars are within the range $10<V<21.5$ mag for Rup-1 and  $11<V<21$ mag for Rup-171. 

To derive reliable astrophysical parameters from the {\it UBV}-based analyses, first, we derived the faint magnitude limit of the $V$-band. The distributions of the number of stars versus $V$ magnitudes with 1 mag intervals were constructed (for each cluster) and are presented in the left panels of Fig.~\ref{fig:histograms}. It can be seen from Fig.~\ref{fig:histograms} that the number of stars increases up to $V=19$ mag and decreases after this limit. We concluded that the $V=19$ mag is the faint magnitude limit for both clusters. We used stars brighter than $V=19$ mag in further {\it UBV}-based analyses.

The number of stars within the ranges $17<V\leq18$ and $18<V\leq19$ mag is 79 and 88 (see Table~\ref{tab:photometric_errors}), respectively, for Rup-171. Although the first decrease in the number of stars appears at $V=18$ mag, as seen in Fig.~\ref{fig:histograms}-c, the number of stars for these two ranges is very close to each other. Therefore, we chose $V=19$ mag as the faint magnitude limit for Rup-171. 

The photometric uncertainties adopted as internal errors were those derived from PSF photometry. We calculated mean photometric errors of the $V$ magnitudes, $U-B$, and $B-V$ color indices as functions of $V$ interval magnitudes. These are listed in the upper rows of Table~\ref{tab:photometric_errors} for the two clusters. $V$-band errors at the faint magnitude limit ($V=19$) are 0.022 mag for Rup-1 and 0.043 mag for Rup-171. The mean errors reach up to 0.085 and 0.031 mag in $U-B$ and $B-V$ measurements for Rup-1 at $V=19$ mag, respectively. These values correspond to 0.191 and 0.098 mag for Rup-171.

\subsection{Gaia Astrometric and Photometric Data}
To perform membership analyses, derive visual extension, age, and distance as well as the kinematic properties of Rup-1 and Rup-171, we used the third data release of the {\it Gaia} \citep[{\it Gaia} DR3,][]{Gaia21} astrometric and photometric data. {\it Gaia} DR3 complements the early third data release of {\it Gaia} \citep[Gaia EDR3,][]{Gaia21}, containing 585 million sources with five-parameter astrometric measurements such as equatorial coordinates ($\alpha$, $\delta$), proper-motion components ($\mu_{\alpha}\cos\delta$, $\mu_{\delta}$), and trigonometric parallaxes ($\varpi$) up to $G=21$ mag. New data in {\it Gaia} DR3 includes new estimates of mean radial velocities to a fainter limiting magnitude of $G\sim~14$ mag. The {\it Gaia} photometry presents three optical pass bands of $G$, $G_{\rm BP}$ and $G_{\rm RP}$ with 330-1950 nm, 330-680 nm, and 630-1050 nm wavelengths, respectively \citep{Evans18}. 

In the study, we gathered {\it Gaia} DR3 astrometric, photometric, and spectroscopic data for all stars in the directions of the studied clusters for $25\times25$ arcmin regions about the clusters' centers. The central locations were taken from \citet{Cantat-Gaudin20} ($\alpha=06^{\rm h} 36^{\rm m} 20^{\rm s}\!\!.\,16$, $\delta= -14^{\circ} 09^{\rm '} 25^{\rm''}\!\!.\,20$ for Rup-1 and $\alpha=18^{\rm h} 32^{\rm m} 02^{\rm s}\!.\,87$, $\delta= -16^{\rm o} 03^{\rm '} 43^{\rm''}\!\!.\,20$ for Rup-171).  The identification charts of the 25 arcmin fields of view for the two clusters are shown in Fig.~\ref{fig:ID_charts}. The final {\it Gaia} catalog includes 21,149 and 362,080 stars within the $8<G<23$ and $7<G<23$ mag ranges for Rup-1 and Rup-171, respectively. When considering these counts it is worth remembering that Rup-171 is located along the Galactic plane. 

To obtain precise results also in the {\it Gaia}-based analyses, we determined the faint magnitude limit of $G$-band through a similar approach as for the $V$-band magnitudes. We plotted histograms with 0.5 bin intervals of $G$ and found that the number of stars decreases after $G=20.5$ mag for the two clusters (see the right-hand panels of Fig.~\ref{fig:histograms}). Hence, we considered this limit as a faint $G$ magnitude limit and used the stars brighter than $G=20.5$ mag for further analyses. Additionally, to visualize our observational field of view with the field of $25'\times25'$ in $G$ bands, we constructed the histogram of stars detected in {\it UBV} bands (red histograms on the right-hand panels of Fig.~\ref{fig:histograms}). Because of a lack of observational data in our field of view we considered $G=20.5$ mag as the limiting magnitude for the {\it Gaia}-based analyses. The mean photometric errors were calculated (for the $25'\times25'$ cluster regions), considering internal errors of $G$ magnitudes, $G_{\rm BP}-G_{\rm RP}$ and $G-G_{\rm RP}$ colors as a function of interval $G$ magnitude. The mean errors for {\it Gaia} photometry are listed in the bottom panel of Table~\ref{tab:photometric_errors}. The mean $G$ errors reach up to 0.012 mag and 0.016 mag, and $G_{\rm BP}-G_{\rm RP}$ errors do not exceed 0.24 mag and 0.35 mag for the stars brighter than $G=21$ mag (which contains faint $G$ limit) for Rup-1 and Rup-171, respectively. The mean $G-G_{\rm RP}$ errors are 0.109 mag and 0.158 mag for relevant $G$ ranges for Rup-1 and Rup-171, respectively. 

\begin{table*}
	\setlength{\tabcolsep}{12pt}
	\renewcommand{\arraystretch}{1.0}
	\fontsize{8pt}{10pt}\selectfont
	\centering
	\caption{\label{tab:photometric_errors}
		{The mean internal photometric errors and number of measured stars in the corresponding $V$ apparent-magnitude interval for each cluster.}}
	\begin{tabular}{ccccc|ccccc} 
		\hline \\[-2ex]
		\multicolumn{5}{c}{Rup-1} & \multicolumn{4}{c}{Rup-171} \\ 
		\hline 
		\rule{0pt}{1ex} $V$ & $N$ & $\sigma_{\rm V}$ & $\sigma_{\rm U-B}$ & $\sigma_{\rm B-V}$ & $N$ & $\sigma_{\rm V}$ & $\sigma_{\rm U-B}$ & $\sigma_{\rm B-V}$ \\[0.5ex]
		\hline 
		\rule{0pt}{1.5ex}
		(8, 12]  &  5 & 0.095 & 0.115 & 0.135 &  2 & 0.008 & 0.015 & 0.011 \\
		(12, 14] &  9 & 0.037 & 0.045 & 0.051 &  6 & 0.006 & 0.012 & 0.011 \\
		(14, 15] & 11 & 0.015 & 0.019 & 0.022 & 36 & 0.010 & 0.014 & 0.016 \\
		(15, 16] & 11 & 0.007 & 0.011 & 0.009 & 62 & 0.010 & 0.023 & 0.018 \\
		(16, 17] & 18 & 0.011 & 0.019 & 0.015 & 42 & 0.012 & 0.033 & 0.023 \\
		(17, 18] & 19 & 0.013 & 0.042 & 0.020 & 79 & 0.024 & 0.087 & 0.051 \\
		(18, 19] & 43 & 0.022 & 0.085 & 0.031 & 88 & 0.043 & 0.191 & 0.098 \\
		(19, 20] & 37 & 0.043 & 0.098 & 0.073 & 52 & 0.090 & 0.272 & 0.192 \\
		(20, 22] & 33 & 0.084 & 0.195 & 0.142 &  3 & 0.152 &  ---  & 0.255 \\ 
		\hline \\
		\multicolumn{5}{c}{Rup-1}  & \multicolumn{4}{c}{Rup-171} \\
		\hline
		\rule{0pt}{1ex}$G$ & $N$ & $\sigma_{\rm G}$ &  $\sigma_{G_{\rm BP}-G_{\rm RP}}$ & $\sigma_{\rm G-G_{\rm RP}}$ &  $N$  & $\sigma_{\rm G}$ & $\sigma_{G_{\rm BP}-G_{\rm RP}}$&$\sigma_{\rm G-G_{\rm RP}}$ &\\[0.5ex]
		\hline
		\rule{0pt}{1.5ex}
		(5, 10]  & 10   & 0.003 & 0.006 & 0.005   & 24    & 0.003 & 0.010  &0.007   &  \\
		(10, 12] & 66        & 0.003 & 0.007 & 0.006   & 148   & 0.003 & 0.007  &0.006   &  \\
		(12, 13] & 95        & 0.003 & 0.005 & 0.005   & 332   & 0.003 & 0.009  &0.007   &  \\
		(13, 14] & 223       & 0.003 & 0.005 & 0.005   & 925   & 0.003 & 0.010  &0.007   &  \\
		(14, 15] & 442       & 0.003 & 0.006 & 0.005   & 2507  & 0.003 & 0.009  &0.007  &  \\
		(15, 16] & 829       & 0.003 & 0.007 & 0.006   & 5679  & 0.003 & 0.011  &0.007   &  \\
		(16, 17] & 1383      & 0.003 & 0.010 & 0.007   & 12220 & 0.003 & 0.016  &0.008   &  \\
		(17, 18] & 2396      & 0.003 & 0.020 & 0.010   & 22331 & 0.004 & 0.031  &0.014   &  \\
		(18, 19] & 3812      & 0.004 & 0.045 & 0.020   & 47034 & 0.005 & 0.066  &0.026   &  \\
		(19, 20] & 5140      & 0.005 & 0.095 & 0.038   & 110109& 0.008 & 0.156  &0.060   &  \\
		(20, 21] & 6461      & 0.012 & 0.239 & 0.109   & 159186& 0.016 & 0.352  &0.158   &  \\
		(21, 23] & 292       & 0.029 & 0.467 & 0.223   & 1585  & 0.033 & 0.552  &0.295   &  \\
		\hline
	\end{tabular}%
\end{table*}%

\subsection{Structural Parameters of the Clusters}
\label{section:rdps}
\begin{figure}
	\centering
	\includegraphics[width=0.98\linewidth]{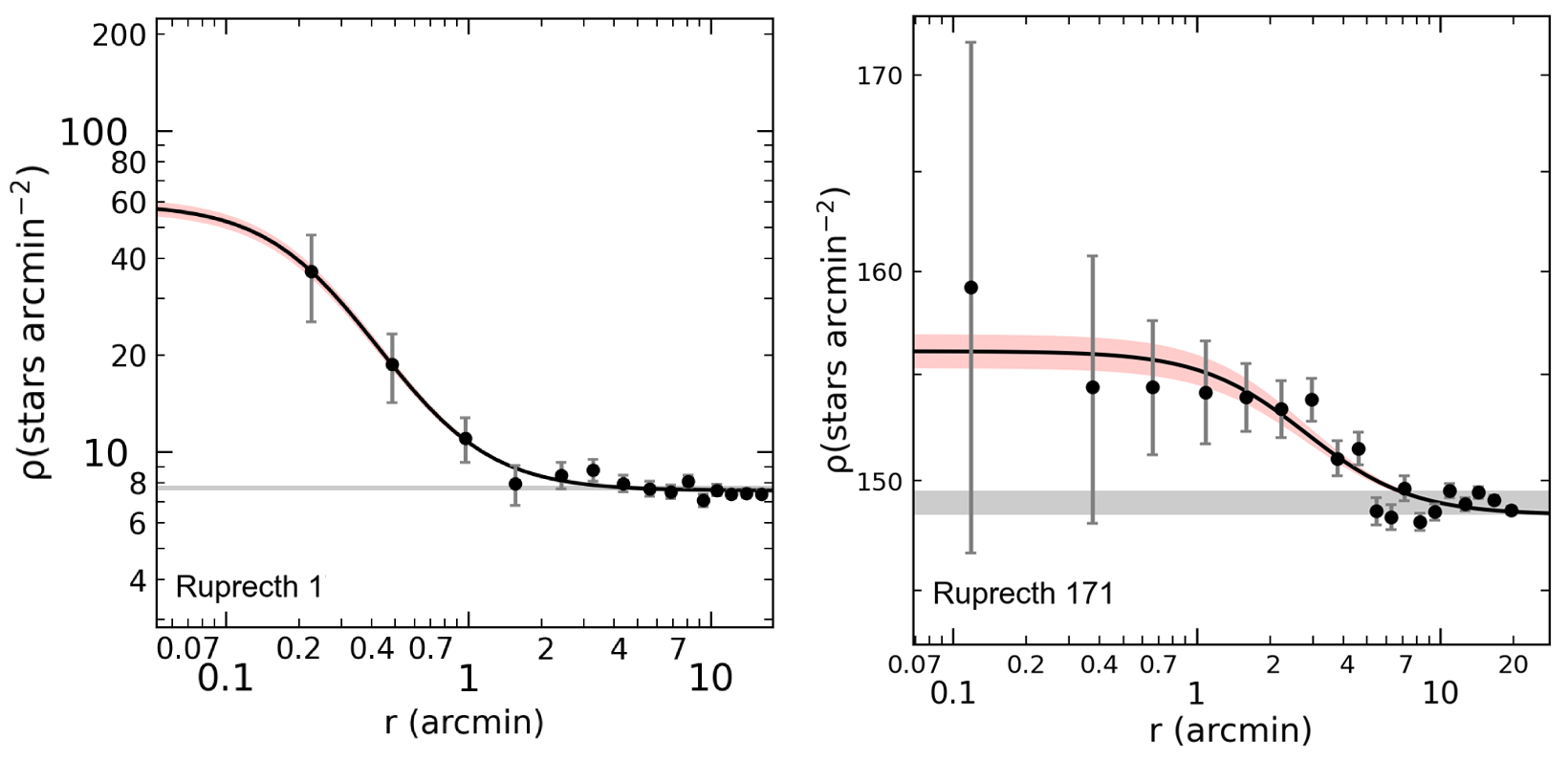}\\
	\caption{Radial Density Profiles for Rup-1 and Rup-171. The best fit \citet{King62} model is represented with a black-continuous line. The horizontal gray band depicts the background density level and its errors, while the red shaded area shows the $1\sigma$ uncertainty of model fit. Stellar density errors were estimated via $1/\sqrt N$.} 
	\label{fig:king}
\end{figure}

Estimation of the structural parameters and visual sizes for the two clusters was based on {\it Gaia} DR3 data of an $25'\times25'$ area centered on each of the clusters. To do this, we utilized radial density profile (RDP) analyses, taking into account the central coordinates presented by \citet{Cantat-Gaudin20}. We divided the cluster areas into concentric rings, each representing a specific distance from the clusters' adopted center. The number of stars within each ring was then counted, and the stellar densities ($\rho$) were computed by dividing the star count by the ring's area. We plotted stellar densities according to distance from the cluster center as shown in Fig.~\ref{fig:king}. We compared RDPs by fitting \citet{King62} models via least-square fitting ($\chi^2$). This allowed us to infer `optimal' estimates for the core, limiting, and effective radii for two clusters. The \citet{King62} model is expressed as $\rho(r)=f_{\rm bg}+[f_{\rm 0}/(1+(r/r_{\rm c})^2)] $, where $r$ is the radius from the cluster center, $f_{\rm bg}$ the background density, $f_{\rm 0}$ the central density, and $r_{\rm c}$ the core radius. The best fitting solution of the \citet{King62} RDP fits for each cluster was represented by a black continuous line in Fig.~\ref{fig:king}. The estimates of central stellar density, core radius, and background stellar density are $f_{\rm 0}=51.550\pm 3.132$ stars arcmin$^{-2}$, $r_{\rm c}=0.254\pm 0.016$ arcmin and $f_{\rm bg}=7.573\pm 0.136$ stars arcmin$^{-2}$ for Rup-1, respectively, and $f_{\rm 0}=7.610\pm 0.973$ stars arcmin$^{-2}$, $r_{\rm c}=3.297\pm 0.920$ arcmin and $f_{\rm bg}=148.411\pm 2.487$ stars arcmin$^{-2}$ for Rup-171, respectively. Through visual examination of the RDP plots, we estimated the observable limiting radii for the two clusters. We adopted these radii as the point where the background density merges with the cluster density. Following this process we estimated the limiting radii as $r=7'$ for Rup-1 and $r=10'$ for Rup-171. Only stars within these limiting radii were included in the following {\it Gaia}-based analyses. 

\begin{figure*}[ht!]
	\centering
	\includegraphics[width=10cm]{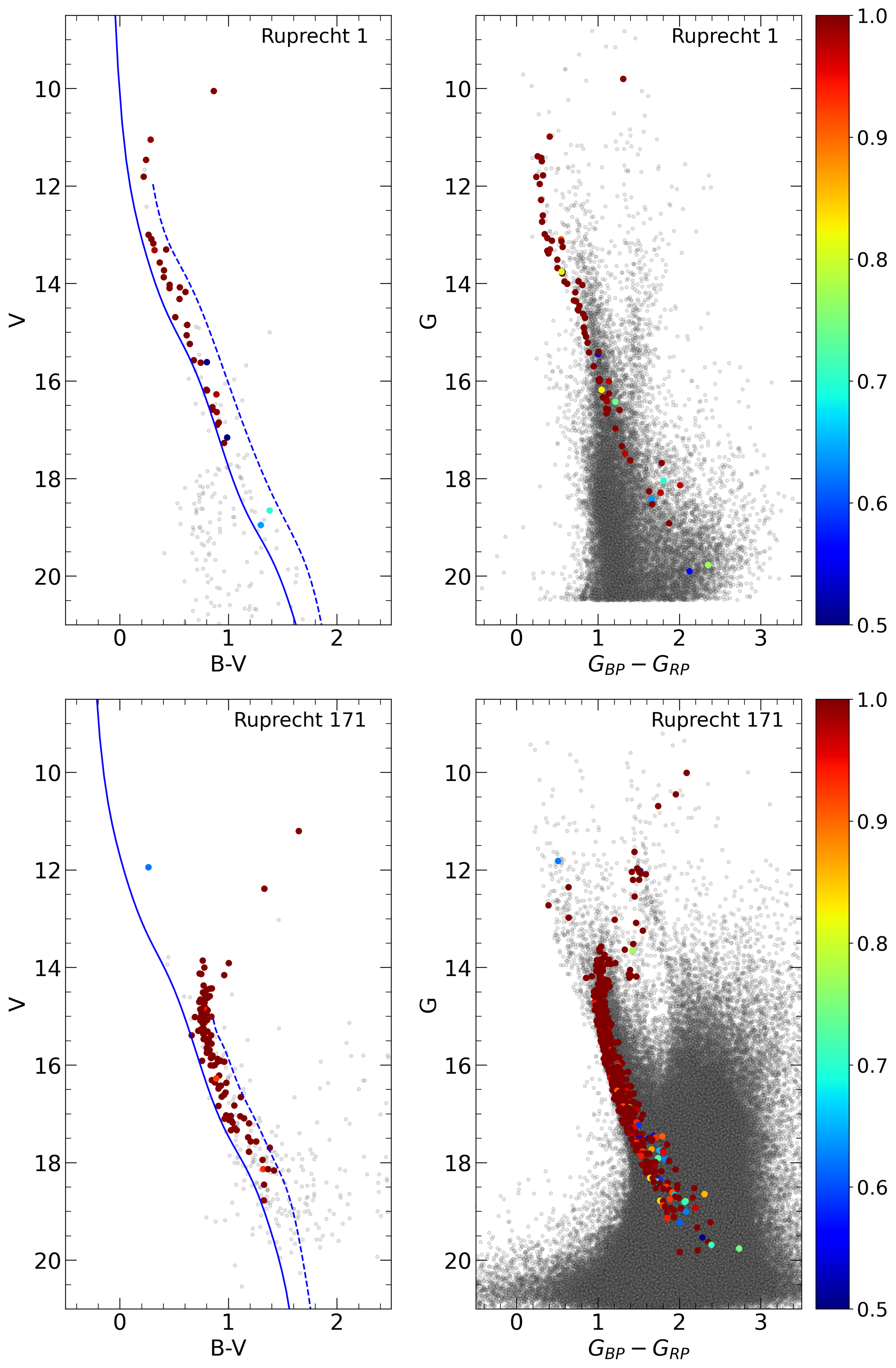}
	\caption{{\it UBV} and {\it Gaia} photometry-based CMDs for the clusters Rup-1 (upper panels) and Rup-171 (lower panels). The blue continuous and dashed lines represent the blue and red envelope of the zero-age main sequence \citep{Sung13}. Color-scaled points show the most probable members with $P\geq0.5$, whereas gray points indicate the stars with membership probabilities $P<0.5$. Most probable member ($P\geq0.5$) stars in the right panels are located within the $r_{\rm lim}\leq7'$ (Rup-1) and $r_{\rm lim}\leq10'$ (Rup-171) radii calculated from {\it Gaia} catalogs, respectively. Gray points in these diagrams represent the low-probability stars ($P<0.5$) located outside the clusters' radii.} 
	\label{fig:cmds}
\end{figure*}

\subsection{Color-Magnitude Diagrams and Selection of Cluster Members}\label{section:cmds}	
Field star contamination across our view of an OC affects the reliable estimation of fundamental parameters for the cluster. It is therefore necessary to separate cluster members from field stars. Thanks to the {\it Gaia} DR3 astrometric data, membership determination analyses give precise results. This leads to the cluster morphology being clearly distinguished on CMDs, allowing precise determinations of the parameters. In this study, we used the Unsupervised Photometric Membership Assignment in Stellar Cluster program \citep[{\sc upmask};][]{Krone-Martins14} method to investigate the membership probabilities of stars in each cluster region. {\sc upmask} is based on the principle that cluster stars share common features in proper-motion and trigonometric parallax space and have a region of concentration in equatorial coordinates. This method was previously used in many studies \citep{Cantat-Gaudin20, Castro-Ginard20, Banks20, Akbulut21, Koc22, Tasdemir23,Yontan23b}. A detailed description can be found in \citet{Cantat-Gaudin18}. 

In the membership analyses, we used equatorial coordinates ($\alpha$, $\delta$), as well as the {\it Gaia} DR3 proper-motion components ($\mu_{\alpha}\cos\delta$, $\mu_{\delta}$) and trigonometric parallaxes ($\varpi$) with their uncertainties as input parameters for all stars in the 25 arcmin regions of the Rup-1 and Rup-171 OCs. We ran 100 iterations of {\sc upmask} for the two clusters, scaling these inputs to unit variance to determine membership probabilities ($P$). We considered the stars with membership probabilities over 0.5 as the most probable cluster members. Hence, for Rup-1 we identified that 74 possible members, brighter than $G=20.5$ mag, lie within the limiting radius ($r\leq7'$) and with membership probabilities $P\geq 0.5$. According to a similar magnitude limit ($G\leq20.5$) and membership criteria ($P\geq 0.5$) with a $r\leq10'$ limiting radius, we identified 596 possible members for Rup-171. \citet{Cantat-Gaudin_Anders20} used {\it Gaia} DR2 data and determined 129 and 739 member stars with membership probabilities over than 0.5 for Rup-1 and Rup-171, respectively. The  {\it Gaia} DR3 data used for membership analyses in this study contain improved precision in position, trigonometric parallax, and proper motion measurements. This improved data could affect the membership analyses. In addition to the membership probabilities $P \geq 0.5$, we considered the stars within the clusters' limiting radii as possible members. These features can explain the differences of the number of member stars between this study and \citet{Cantat-Gaudin_Anders20}. We used these stars in further analyses for the determination of mean astrometric and kinematic parameters, as well as the ages and distances of the two clusters. $G\times (G_{\rm BP}-G_{\rm RP}$) CMDs of these stars within the aforementioned 25 arcmin fields are shown in the upper and lower panels of Fig.~\ref{fig:cmds} for Rup-1 and Rup-171, respectively. In order to perform {\it UBV} photometry-based analyses for the two clusters, the membership probability values calculated from the {\it Gaia} catalog were also applied to the same stars identified in the {\it UBV} catalog. For this purpose, the stars in the {\it Gaia} and {\it UBV} catalogs were cross-matched according to their coordinates so that the membership probabilities of the same stars in the {\it UBV} catalog were determined. 
\begin{figure*}[h!]
	\centering
	\includegraphics[width=1\linewidth]{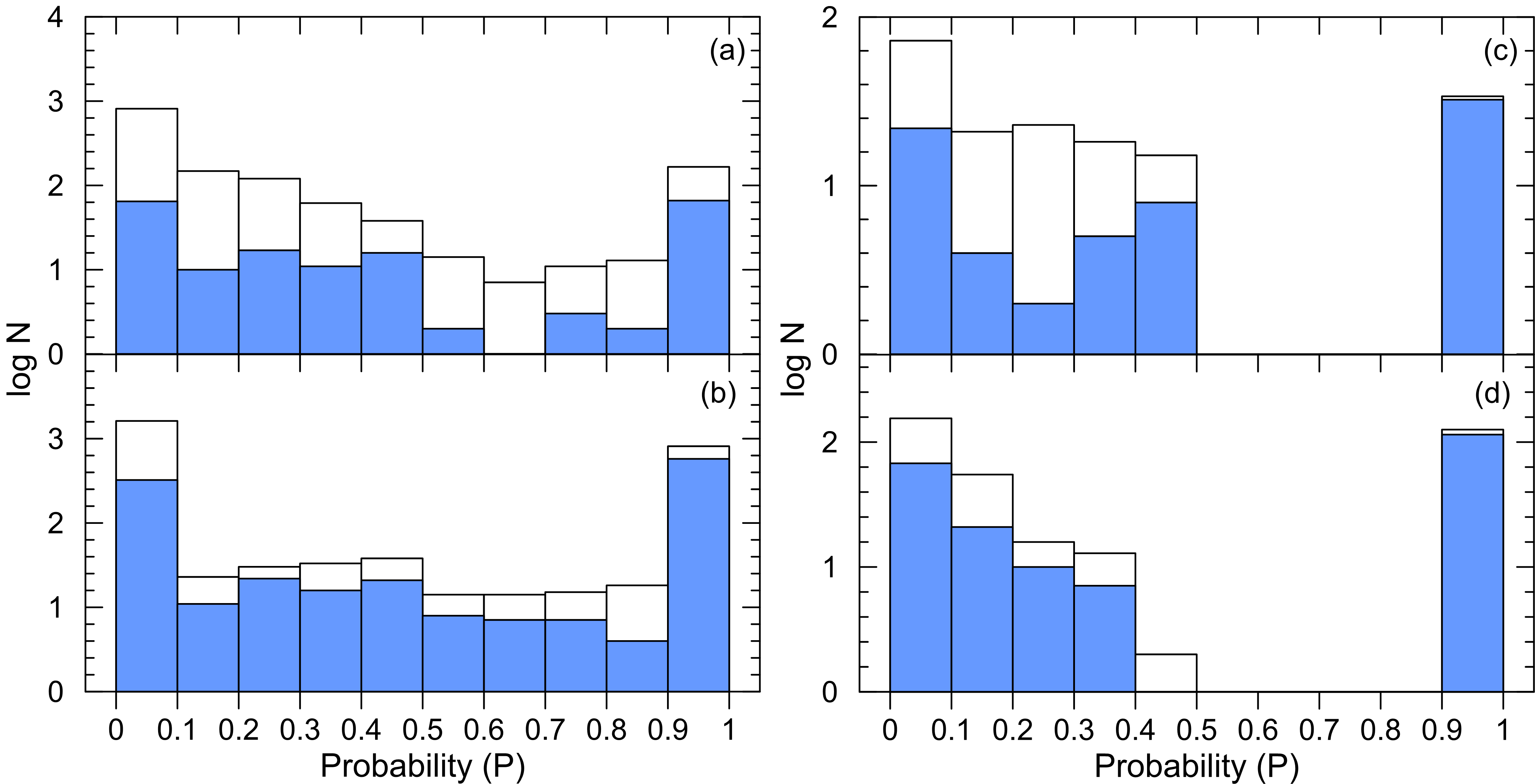}
\caption{Histograms of the membership probabilities versus number of stars for Rup-1 (a) and Rup-171 (b) constructed from {\it Gaia} catalog and for Rup-1 (c) and Rup-171 (d) created from {\it UBV} catalog}. The blue-colored histograms indicate stars within the clusters' limiting radii.
	\label{fig:prob_hists}
\end{figure*}

Additionally, using the photometric criteria for the {\it UBV} data, we took into consideration the possible binary star contamination on the main-sequences of Rup-1 and Rup-171. We plotted the $V\times (B-V)$ CMDs and fitted the Zero Age Main-Sequence (ZAMS) of \citet{Sung13} as a blue and red envelope to these diagrams (see Fig.~\ref{fig:cmds}). The blue envelope of ZAMS was fitted through visual inspection, considering the most probable ($P \geq 0.5$) member stars in the main sequence. For the red envelope ZAMS, the blue one was shifted by 0.75 mag towards brighter magnitudes to include the possible binary star contamination. Through this investigation, 36 and 115 stars remained as the most probable cluster members for {\it UBV} data in Rup-1 and Rup-171, respectively. These stars were used in further estimation of color excess, photometric metallicity as well as the derivation of {\it UBV} data-based age and isochrones distance for each cluster. $V\times (B-V)$ CMDs with the blue and red ZAMS envelopes, as well as the most probable and field stars, are shown in left panels of Fig.~\ref{fig:cmds} for Rup-1 and Rup-171.
\begin{figure*}[h!]
	\centering
	\includegraphics[width=0.85\linewidth]{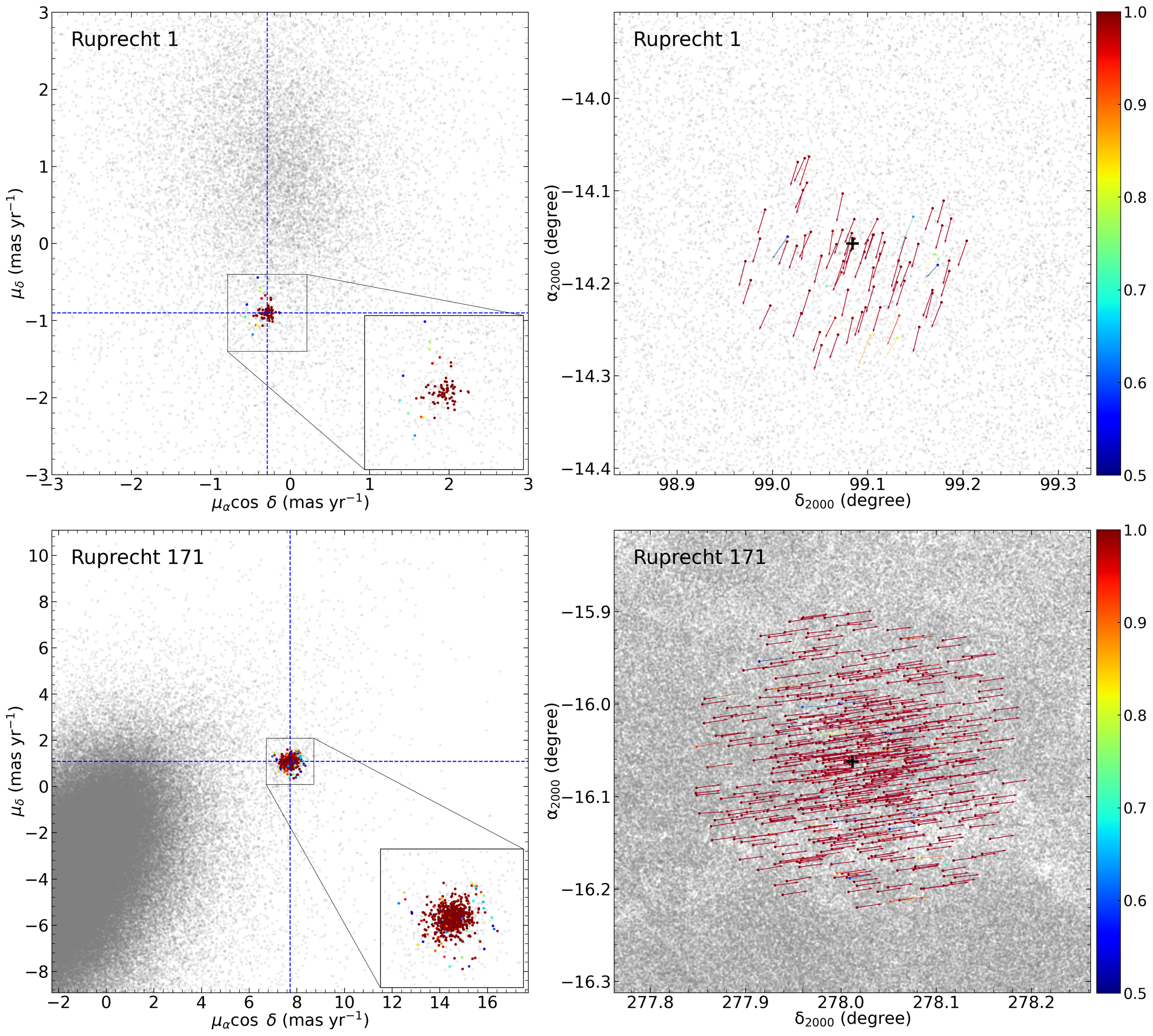}\\
	\caption{Vector-Point Diagrams (VPDs, left panels) and proper-motion components vectors on the equatorial coordinates (right panels) of Rup-1 and Rup-171. Color-scaled points and vectors represent the membership probabilities of the stars with  $P\geq0.5$ according to the color scale shown on the right of the sub-figures. The zoomed-in boxes in the left panels show the areas of concentrated member stars in the VPDs and blue dashed lines are the intersection of the mean proper-motion values. Also in the right panels, black crosshairs indicate the equatorial coordinate centers of the OCs and the magnitude of the vectors is arbitrarily chosen to enable their direction to be seen.}
	\label{fig:vpds} 
\end{figure*}

\begin{figure}[h!]
	\centering
	\includegraphics[width=0.99\linewidth]{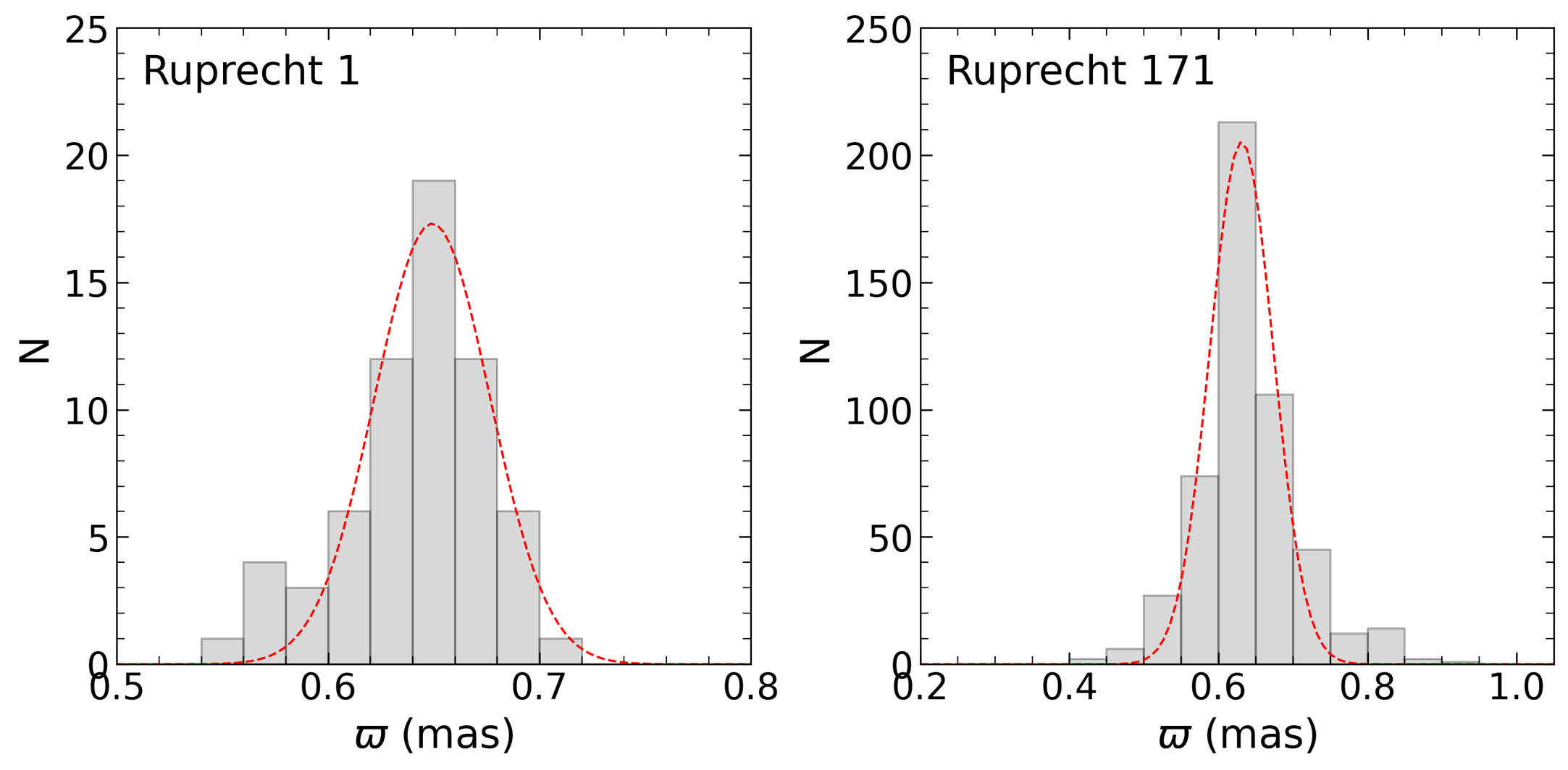}\\
	\caption{Histograms of trigonometric parallax for Rup-1 and Rup-171. Figures are constructed from the most probable member stars located within the limiting radii of the two clusters. Red dashed lines represent the Gaussian fits to the distributions.}
	\label{fig:plx}
\end{figure}

Using membership probabilities and numbers of stars from the {\it Gaia} and {\it UBV} catalogues of each cluster, we prepared probability distributions as shown in Fig.~\ref{fig:prob_hists}. These figures compare the membership probabilities versus number of stars. Panels (a) and (b) in the figures were constructed for each 25-arcmin cluster region (white histograms) and stars inside the clusters' limiting radii (blue histograms), whereas panels (c) and (d) were plotted for the stars detected in {\it UBV} observations (white histograms) as well as lying within the ZAMS curves and clusters' limiting radii (blue histograms). It can be seen from the right panels of the Fig.~\ref{fig:prob_hists} that the membership probability of cross-matched stars in {\it UBV} catalogues are higher than 0.9. These stars were also used to obtain mean proper-motion components and trigonometric parallaxes of both clusters. To assign the member stars in proper-motion space and investigate the bulk motion of the clusters we plotted both vector-point diagrams (VPDs) and projection of proper-motion vectors on the sky, which are presented as left and right panels of Fig.~\ref{fig:vpds}, respectively. In both of the left panels of Fig.~\ref{fig:vpds} it can be seen that the most probable members (the color-scaled points) are concentrated in certain areas, allowing cluster stars to be distinguished from field stars (gray points). The right panels of Fig.~\ref{fig:vpds} indicate that most probable members of the cluster have similar directions on the RA and DEC plane. The mean proper motion component estimates are ($\mu_{\alpha}\cos \delta$, $\mu_{\delta})=(-0.287 \pm 0.003, -0.903 \pm 0.003$) for Rup-1 and ($\mu_{\alpha}\cos \delta$, $\mu_{\delta}) = (7.720 \pm 0.002, 1.082 \pm 0.002$) mas yr$^{-1}$ for Rup-171. The intersections of the blue dashed lines in Fig.~\ref{fig:vpds} show the mean value points of the proper-motion components. Trigonometric parallaxes of the most probable member stars were used to calculate mean trigonometric parallaxes and so corresponding distances of the clusters. To perform these analyzes we constructed the histograms of trigonometric parallaxes versus stellar numbers and fitted a Gaussian to these distributions, as shown in Fig.~\ref{fig:plx}. These distributions include the most probable members with probabilities $P\geq0.5$ and those inside the limiting radii of clusters. From the Gaussian fits these groupings, we obtained the mean trigonometric parallaxes of Rup-1 and Rup-171 as $\varpi= 0.649\pm 0.027$ mas and $\varpi= 0.631 \pm 0.042$ mas, with corresponding distances $d_{\varpi}=1541\pm64$, $d_{\varpi}=1585\pm106$ pc respectively. The mean trigonometric parallax error was calculated from the statistical uncertainties in the Gaussian fitting process.

\begin{figure}[h!]
	\centering
	\includegraphics[width=0.87\linewidth]{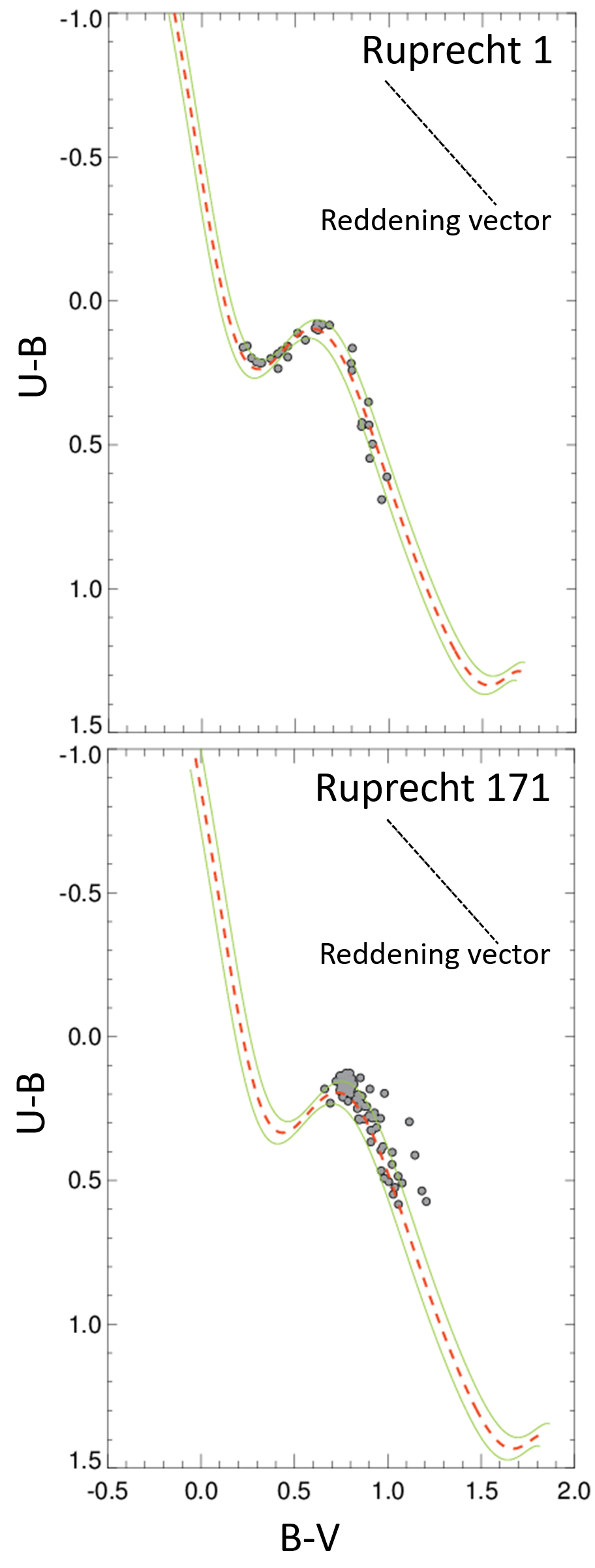}
	\caption{Two-Color Diagrams (TCDs) of the most probable member main-sequence stars of Rup-1 and Rup-171. Red dashed and green solid lines indicate the shifted zero age main sequence (ZAMS) of \citet{Sung13} and $\pm1\sigma$ standard deviations, respectively.
	}\label{fig:tcds} 
\end{figure}

\begin{figure}[h!]
	\centering
	\includegraphics[width=0.95\linewidth]{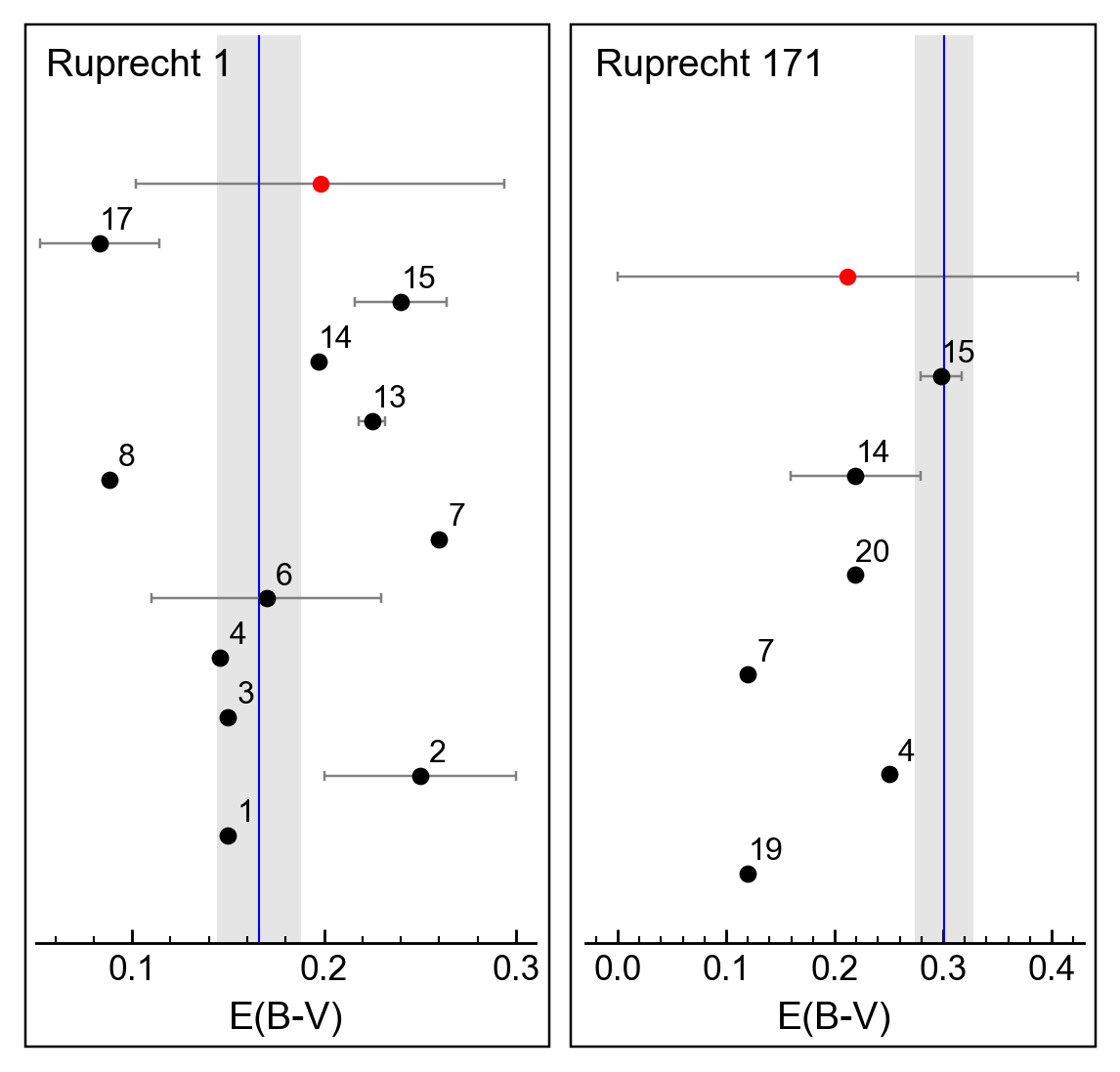}
	\caption{Comparison of the color excess estimated for two OCs in this study with ones given in the literature. Black and red dots indicated by numbers represent the literature data in Table 1 and the color excess calculated from the 3D reddening maps, the blue lines and grey regions show the $E(B-V)$ color excess and uncertainties calculated in this study, respectively.
    }\label{fig:Literature_E(B-V)} 
\end{figure}

\section{Analyses of the {\it UBV} Data}
This section summarizes the procedures for the astrophysical analyses of Rup-1 and Rup-171. We used two-color diagrams (TCDs) to calculate the reddening and photometric metallicities separately. Keeping these two parameters as constants and using CMDs, we next obtained the distances and ages simultaneously \citep[as performed in previous studies such as ][]{Bilir06, Bilir10, Bilir16, Bostanci15, Yontan23b, Yontan23c, Gokmen23}. Hence, we summarized the relevant analyses in this section.

\subsection{Color Excess for the Two Open Clusters}
To obtain the $E(U-B)$ and $E(B-V)$ color excesses in the direction of the two clusters we constructed $(U-B)\times (B-V)$ TCDs. These are shown as Fig.~\ref{fig:tcds} and are based on the most probable ($P\geq 0.5$) main-sequence stars. The intrinsic ZAMS of \citet{Sung13} for solar metallicity was fitted to the observational data, employing the equation of $E(U-B)=0.72 \times E(B-V) + 0.05\times E(B-V)^2$ \citep{Garcia88}. This process was performed according to a least-square ($\chi^2$) method with steps of 0.001 mag. By comparing the ZAMS to the most probable main-sequence stars, we achieved best-fit results for color excesses corresponding to the minimum $\chi^2$. These estimates are $E(B-V)=0.166\pm 0.022$ mag for Rup-1 and $E(B-V)=0.301\pm 0.027$ mag for Rup-171. The errors of the calculations were determined as $\pm 1\sigma$ deviations, and are shown as the green lines in Fig.~\ref{fig:tcds}. Using the equation of $A_{V}/E(B-V)=3.1$ \citep{Cardelli89}, we calculated the $V$-band absorption as $A_{V}=0.511\pm 0.068$ and $A_{V}=0.933\pm 0.083$ mag for Rup-1 and Rup-171, respectively. 

In the study, three-dimensional (3D) reddening maps known as STructuring by Inversion the Local Interstellar Medium ({\sc Stilism})\footnote{https://stilism.obspm.fr/} were used to determine the color excess in the direction of two OCs. We used the 3D reddening map of \citet{Lallement2014}, which analyses stars within 2.5 kpc at about 23,000 sightlines. Using {\sc Stilism} information, considering the Galactic coordinates of two OCs ($l, b$) and their mean distances calculated from trigonometric parallax measurements ($d_{\varpi}$), the color excesses for Rup-1 and Rup-171 were estimated as $E(B-V)=0.198\pm0.096$ and as $E(B-V)=0.212\pm0.212$ mag, respectively.

The comparison of the color excess estimated from the {\it UBV} photometric data for two OCs with the results in the literature (which are given in Table~\ref{tab:literature}) is shown in Fig. \ref{fig:Literature_E(B-V)}. In the panels of the Fig.~\ref{fig:Literature_E(B-V)}, the black dots labelled with numbers represent $E(B-V)$ color excesses given in the literature as identified in Table~\ref{fig:Literature_E(B-V)}, the red dots represent the color excess calculated from the 3D reddening maps, and the blue lines and grey regions represent the $E(B-V)$ color excess and uncertainties calculated in this study. The color excess estimated for Rup-1 is in good agreement with the values ($0.146 \leq E(B-V) \leq 0.197$ mag) estimated by different authors \citep{Kharchenko05, Kharchenko09, Kharchenko13, Oralhan15, Cantat-Gaudin20}. For Rup-171 the estimated color excess is compatible with the results of \citet{Kharchenko13} and \citet{Dias21}. In general, the $E(B-V)$ color excess calculated for Rup-1 in this study are in good agreement with the results in the literature, while the color excess estimated for Rup-171 is close to the upper limit in the literature (see Fig. \ref{fig:Literature_E(B-V)}).

\begin{figure}[b!]
	\centering 
	\includegraphics[width=0.99\linewidth]{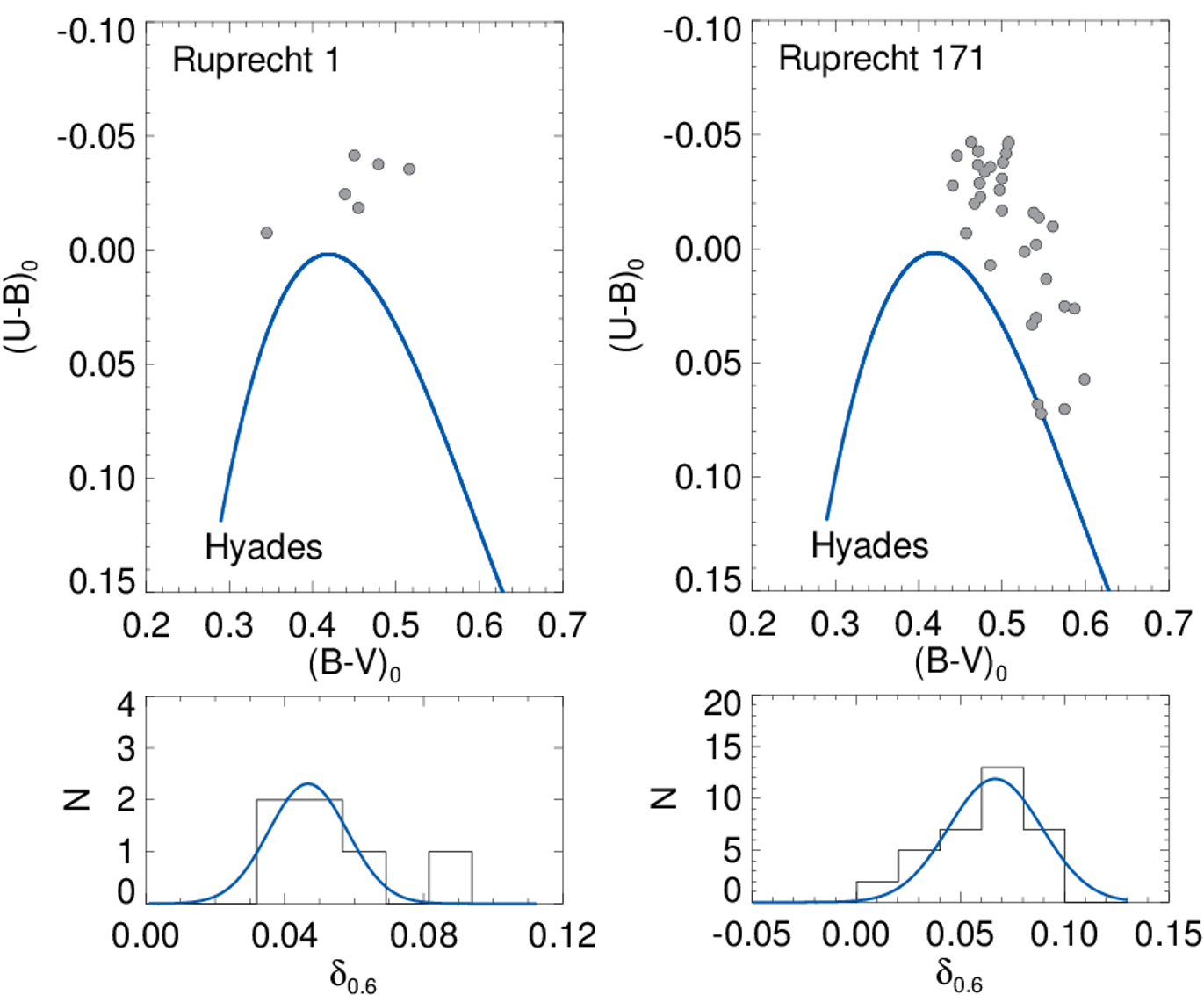}
	\caption{Two-Color Diagrams (TCDs,upper panels) and the histogram of normalised $\delta_{0.6}$ for Rup-1 and Rup-171 (lower left and right panels, respectively). The solid blue lines in TCDs and histograms are the main sequence of Hyades and Gaussian fit, respectively.} \label{fig:hyades}
\end{figure}

\subsection{Photometric Metallicities for the Two Open Clusters}
We estimated the photometric metallicity of the studied clusters using the $(U-B)_0\times(B-V)_0$ TCDs and employing the method of \citet{Karaali03a, Karaali03b,Karaali11}. This methodology considers F and G spectral-type main-sequence stars and their UV-excesses. The $(B-V)_0$ color indices of these stars are in the range of $0.3\leq (B-V)_0\leq0.6$ mag \citep{Eker18, Eker20}. Thus, we estimated intrinsic $(B-V)_0$ and $(U-B)_0$ color indices considering the color excesses derived above and selected the most probable F-G spectral type main-sequence stars inside the $0.3\leq (B-V)_0\leq0.6$ mag range. We determined UV-excesses ($\delta$) for the selected stars. This is described as differences between the $(U-B)_0$ color indices of the selected cluster and Hyades main-sequence members with the same intrinsic $(B-V)_0$ color indices. Such differences are defined by the expression of $\delta =(U-B)_{\rm 0,H}-(U-B)_{\rm 0,S}$, where H and S are the Hyades and cluster stars with the same $(B-V)_0$ color indices, respectively. By normalizing the UV-excess of the stars at $(B-V)_0 = 0.6$ mag we estimated the selected stars' normalized UV-excess ($\delta_{0.6}$) values. For each cluster, we constructed the histogram of $\delta_{0.6}$ and fitted a Gaussian to the resulting distribution to derive a mean $\delta_{0.6}$ value. This was then used in the estimation of the photometric metallicity ($[{\rm Fe/H}]$) of the selected cluster. The equation of \citet{Karaali11} used for metallicity calculations is given as follows:   
\begin{equation}
	\begin{split}
		{\rm [Fe/H]}=-14.316(1.919)\delta_{0.6}^2&-3.557(0.285)\delta_{0.6}\\
		&+0.105(0.039).
	\end{split}
\end{equation}	
Six F-G spectral type main-sequence stars in Rup-1 and 32 stars in Rup-171 were selected to derive the [Fe/H] values of these two clusters. $(U-B)_0\times(B-V)_0$ diagrams and distribution of normalized $\delta_{0.6}$ values of selected stars for Rup-1 and Rup-171 are shown in Fig.~\ref{fig:hyades}. The peaks of the Gaussian fits to the normalized UV-excesses are $\delta_{0.6}=0.047\pm0.011$ and $\delta_{0.6}=0.067\pm0.022$ mag for Rup-1 and Rup-171, respectively. The uncertainty of the mean $\delta_{0.6}$ was derived by $\pm1\sigma$ (the standard deviation) of the Gaussian fit. Taking into account the internal errors of the photometric metallicity calibration, the metallicities corresponding to mean $\delta_{0.6}$ values are calculated to be ${\rm [Fe/H]} = -0.09\pm 0.06$ and ${\rm [Fe/H]} = -0.20\pm 0.13$ dex, for Rup-1 and Rup-171, respectively. Moreover, considering uncertainties of the {\it UBV} data and relevant color excesses, we determined external errors as 0.15 dex for both clusters.  We evaluated these two values using error propagation. Hence, the final results for metallicities were determined as ${\rm [Fe/H]} = -0.09 \pm 0.16$ and ${\rm [Fe/H]} = -0.20 \pm 0.20$ dex, for Rup-1 and Rup-171, respectively.   

The calculated metallicities were transformed into the mass fraction $z$ to help select which isochrones would be used in age estimation. We considered the analytic equation given in the studies of \citet{Gokmen23} and \citet{Yontan23b}. The equation is given as follows:

\begin{equation}
	z=\frac{(z_{\rm x}-0.2485\times z_{\rm x})}{(2.78\times z_{\rm x}+1)}
        \vspace{5pt}
\end{equation} 
Here, $z$ and $z_{\rm x}$ indicate the elements heavier than helium, and the intermediate operation function which is expressed by \vspace{3pt}
\begin{equation}
\scalemath{1.2}{%
z_{\rm x}=10^{\large \left [{\rm[ Fe/H]}+\log\large\left(\frac{z_{\odot}}{1-0.248-2.78\times z_{\odot}}\large \right)\large \right]}%
}\vspace{3pt}
\end{equation} 
\noindent
respectively. $z_{\odot}$ is the solar metallicity adopted as 0.0152 \citep{Bressan12}. We calculated $z=0.012\pm 0.003$ for Rup-1 and $z=0.010\pm 0.004$ for Rup-171.

In the literature, the metallicity estimation of Rup-1 is based on the adoption of theoretical metal contents (see Table~\ref{tab:literature} on page~\pageref{tab:literature}). The photometric metallicity calculated in the current study for Rup-1 matches well the value of \citet{Dias21}. \citet{Casamiquela21} analyzed high-resolution {\sc hermes} spectra of six red clumps in Rup-171, measuring the metallicity of the cluster as [Fe/H]=$-0.041\pm0.014$ dex. \citet{Casali20} used the {\sc harps-n} spectrograph \citep{Cosentino14} and acquired high resolution ($R \sim$ 115,000) optical spectra for the eight highly probable members of Rup-171 including two red giant branch (RGB) and six red clump stars (RC). They performed two analysis methods.  The first, Fast Automatic MOOG Analysis ({\sc fama}), is based on the equivalent width method and the second is based on the analysis code {\sc rotfit} \citep{Frasca06, Frasca19}. Hence they reported two different metallicity values for each studied star. According to {\sc fama} and {\sc rotfit} analyses, \citet{Casali20} found that the metallicities of the six clump stars are within the $-0.38\leq\rm[Fe/H]\leq 0.08$ and $-0.12\leq\rm[Fe/H]\leq 0.10$ dex, in order of the two methods. They indicated that RGB stars are more metal-poor than RC stars and there are residual differences between the metallicity values because of the physics of stellar evolution, such as atomic diffusion and mixing or to approaches during the spectroscopic analyses. Hence, they adopted the mean metallicity result from six RC stars derived from {\sc rotfit} analyses as [Fe/H]=$0.09 \pm 0.10$ dex. Our metallicity estimate ($-0.20 \pm 0.20$ dex) for Rup-171 is based on F-G type main-sequence stars, and it is more metal-poor than the literature studies. It's important to note that the increase in metallicities for giant stars is not solely due to a single factor but rather a combination of various processes. Stellar nucleosynthesis, mass loss, and mixing processes with convective motion during the giant stage are more efficient compared to main-sequence stars, leading to a higher enrichment of metals in the outer layers \citep{Bitsch20, Wang23}. For the reasons mentioned above, the use of metallicity calculated from main-sequence stars in OC age calculations may give more precises results.

\subsection{Distance Moduli and Age Estimation}
\label{distance_age}
\begin{figure*}[ht!]
	\centering
	\includegraphics[width=0.85\linewidth]{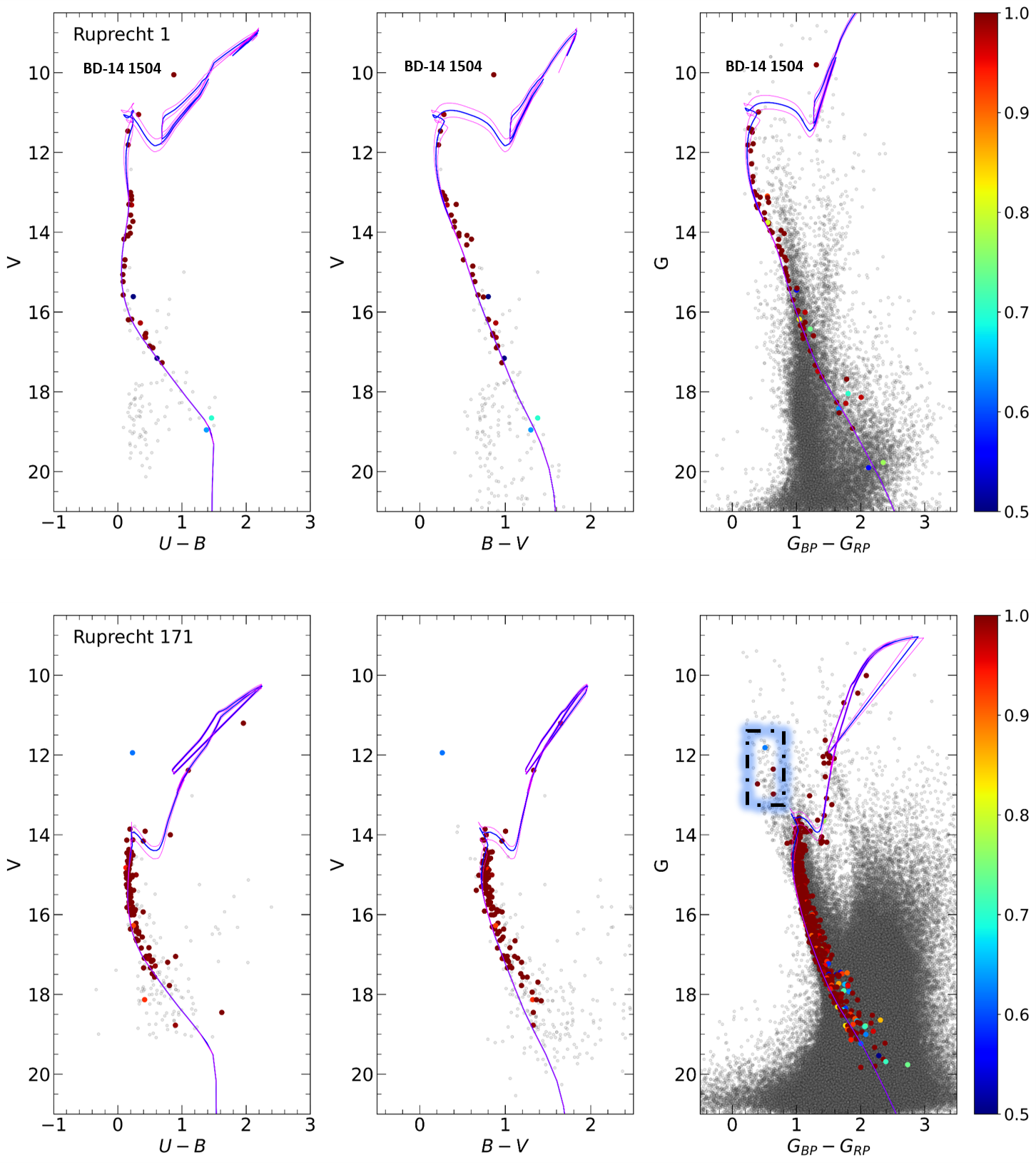}
	\caption{CMDs for Rup-1 (upper panels) and Rup-171 (lower panels). The differently colored dots represent the membership probabilities according to the color scales shown on the right side of the diagrams. Gray dots indicate low-probability members ($P<0.5$), or field stars ($P=0$). The blue lines show the best-fit {\sc parsec} isochrones, while the purple lines are their associated errors. The area bounded by the blue-shaded dash-dotted line contains the BSSs of Rup-171 (lower right panel).
		}\label{fig:age_cmds}
\end{figure*}

The distance moduli, distances, and ages of Rup-1 and Rup-171 were estimated by fitting the {\sc parsec} isochrones of \citet{Bressan12} to the {\it UBV} and {\it Gaia} based CMDs, as shown in Fig.~\ref{fig:age_cmds}. Selection of the {\sc parsec} isochrones was made according to the mass fractions ($z$) derived above for the two clusters. A fitting procedure to the $V\times (U-B)$, $V\times (B-V)$, and $G\times (G_{\rm BP}-G_{\rm RP})$ CMDs was applied by visual inspection taking into account the most probable ($P\geq 0.5$) main-sequence, turn-off, and giant member stars present in the two studied clusters: The first step in the fitting process was to ensure that the isochrones had the best fit to the lower envelope of the most likely main sequence stars. After this step, the isochrones with the ages that best represent the turn-off point of the cluster and the most likely stars in the giant region were determined. While determining the distance moduli and ages from the {\it UBV} data, we used $E(U-B)$ and $E(B-V)$ color excesses obtained in this study. For the {\it Gaia} data, when we were taking into account the coefficient of $E(G_{\rm BP}-G_{\rm RP})= 1.41\times E(B-V)$ as given by \citet{Sun21}, we interpreted that this value does not match well with the observational value to determine distance moduli and age for the two clusters. We achieved a better estimation of these two astrophysical parameters from {\it Gaia} data using the coefficient of $E(G_{\rm BP}-G_{\rm RP})= 1.29\times E(B-V)$ as given by \citet{Wang19}. The errors for the distance moduli were calculated with the method of \citet{Carraro17}. We estimated the uncertainty in the derived cluster ages by fitting two more isochrones whose values were good fits to the data sets but at the higher and lower acceptable values compared to the adopted mean age. The best fit isochrones represent the estimated ages for the clusters, whereas the other two closely fitting isochrones, where one is younger and the other is older than the estimated best fit age, were taken into consideration to estimate the uncertainties in cluster age. Thus, errors for the ages contain visual inspection errors and do not contain errors of the estimated distance moduli, color excesses, and metallicities.

The estimation of the distance modulus, distance, and age parameters for the two clusters are as follows:


\begin{itemize} 
\item{{\it Rup-1}: The best fit by the $z=0.012\pm 0.003$ scaled {\sc parsec} isochrones across $\log(t)$=8.68, 8.76, and 8.83 yr ages gave the apparent distance modulus and age of the cluster as $\mu_{\rm V}=11.346 \pm 0.083$ mag and $t=580 \pm 60$ Myr, respectively. The best age and distance modulus solution in {\it UBV} and {\it Gaia} photometry is shown in the upper panels of Fig.~\ref{fig:age_cmds}. By applying the estimated distance modulus, ($\mu_{\rm V}$), and $V$-band absorption ($A_{\rm V}$) values into the distance modulus definition ($\mu_{\rm V}=5\times \log d-5 + A_{\rm V}$), we calculated the distance of the cluster to be $d_{\rm iso}=1469\pm 57$ pc. The age and distance determined in this study are in reasonable agreement with most of the results given by different researchers (see Table~\ref{tab:literature} on page~\pageref{tab:literature}). The isochrone fitting distance of the cluster agrees within error for the distance calculated above from trigonometric parallax ($d_{\varpi}=1541\pm64$ pc, Sec.~\ref{section:cmds}).
		
The best-fitting isochrone is well-matched with the position of the most probable members except the brightest star on clusters' CMDs (upper panels of Fig.~\ref{fig:age_cmds}). According to information from the SIMBAD database, this star is classified as a double or multiple star named BD-14 1504. In our study, we determined the apparent $V$-band magnitude of BD-14 1504 as $V=10.054$ (which corresponds to $G=9.806$ mag in  {\it Gaia} DR3 data) with the probability of $P=1$. The star is located at a distance of $2^{'}.6$ from the center of the cluster. Moreover, {\it Gaia} DR3 proper motion components ($\mu_{\alpha}\cos \delta, \mu_{\delta}=-0.416 \pm 0.027, -0.937 \pm 0.030$ mas yr$^{-1}$) and trigonometric parallax ($\varpi=0.598\pm0.032$ mas) values were well-matched with the mean results of these parameters of Rup-1. Astrometric evidence indicates that BD-14 1504 is a member of Rup-1. It's important to note that the apparent magnitude of a double or multiple-star system is influenced by the combined magnitudes of its components, the brightness ratio between the components, and the separation between them. These factors can result in variations in the observed magnitudes of the system over time \citep{Minnaert69}. These processes in double or multiple systems potentially can explain why star BD-14 1504 is not superimposed with the age isochrones fitted to the cluster's CMDs in the current paper.}

\item{{\it Rup-171}: The isochrones of $\log(t)$=9.38, 9.43, and 9.48 yr with $z=0.010\pm 0.004$ were fitted on the {\it UBV} and {\it Gaia} based CMDs, as shown in the upper panels of Fig.~\ref{fig:age_cmds}. Based on this isochrone fitting, the distance modulus, distance, and age for Rup-171 are $\mu_{\rm V}=11.819 \pm 0.098$ mag, $d_{\rm iso}=1509\pm 69$ pc, and $t=2700\pm 200$ Myr, respectively. The age and distance values derived for the cluster are also in good agreement with most of the findings presented by earlier studies (see Table~\ref{tab:literature}). The isochrone-based distance estimate also matches within error with the mean trigonometric parallax ($d_{\varpi}=1585\pm106$ pc, Sec.~\ref{section:cmds}) calculated earlier in this study.       
    
We investigated the {\it Gaia}-based CMD of the cluster and picked out the blue straggler stars (BSS) by visual inspection. We identified four BSSs with probabilities over 0.6 within the radial distance of 5$^{'}$ from Rup-171's center. The BSSs of the cluster are plotted in the blue dotted dashed-lined box and shown in the lower right panel of Fig.~\ref{fig:age_cmds}. \citet{Jadhav21} investigated 1246 OCs with {\it Gaia} DR2 data and identified BSSs within these clusters. They classified seven BSSs, four of them with possible candidates of BSSs. The $G$-band magnitude of four BSSs found in this study are within the $11<G<13$ mag range, and they are in common with the stars that confirmed as cluster's BSSs in the study of \citet{Jadhav21}. When the four possible BSSs of \citet{Jadhav21} were examined according to {\it Gaia} DR3 data, it was found that their magnitudes and color indices are within the ranges $14<G<14.5$ and $0.8<(G_{\rm BP}-G_{\rm RP})<1$ mag, respectively, which occur at the most probable MS turn-off point as can be seen in the lower right panel of Fig.~\ref{fig:age_cmds}. Hence, we concluded that these stars are not good BSS candidates.} 
\end{itemize}
\begin{figure*}[b!]
	\centering
	\includegraphics[width=1.\linewidth]{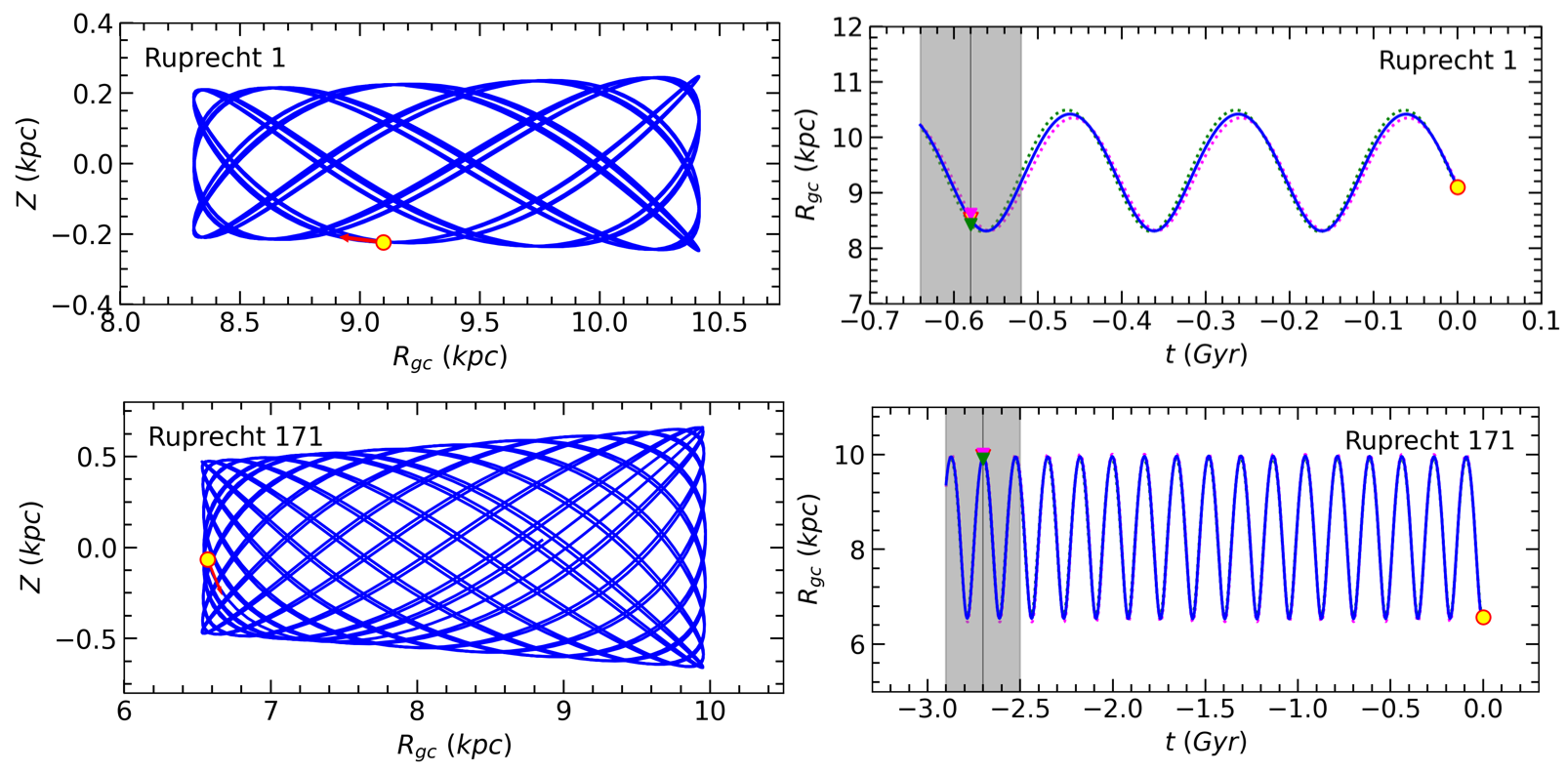}
	\caption{The Galactic orbits of Rup-1 (upper panels) and Rup-171 (lower panels) in the $Z \times R_{\rm gc}$ and $R_{\rm gc} \times t$  planes. The filled yellow triangles and circles are the birth and current locations, respectively. Red arrows show the motion vectors. Pink and green colors with dashed lines and filled triangles indicate the clusters' orbit and birth radius for upper and lower errors of input parameters. The gray shaded regions in right panels show the area of errors in age, and the vertical solid lines present the ages of two OCs corresponding to the birth positions.}
  \label{fig:galactic_orbits} 
\end{figure*} 

\section{Kinematics and Galactic Orbit Parameters of Two Open Clusters}
We estimated kinematical properties and the Galactic orbital parameters of Rup-1 and Rup-171 using the {\sc MWPotential2014} potential model as implemented in {\sc galpy} (the galactic dynamics library) and described by \citet{Bovy15}\footnote{See also https://galpy.readthedocs.io/en/v1.5.0/}. The {\sc MWPotential2014} model is a simplified representation of the Milky Way, assuming axis-symmetry and time-independence of the potential. It consists of a spherical bulge, a dark matter halo, and a Miyamoto-Nagai \citep{Miyamoto75} disk potential. The spherical bulge represents the mass distribution of the Milky Way, and it is defined as a spherical power-law density profile as described by \citet{Bovy15}, given as follows:
\begin{equation}
	\rho (r) = A \left( \frac{r_{\rm 1}}{r} \right) ^{\alpha} \exp \left[-\left(\frac{r}{r_{\rm c}}\right)^2 \right] \label{eq:rho}
\end{equation} 
In this expression, $r_{\rm 1}$ represents the current reference radius, $r_{\rm c}$ the cut-off radius,  $A$ the amplitude that is applied to the potential in mass density units, and $\alpha$ is the power-law index that determines the steepness of the density profile. 

The disk potential describes the gravitational potential of a disk-like structure in Galactic dynamics as described by \citet{Miyamoto75}, given as follows:	
\begin{equation}
	\Phi_{\rm disk} (R_{\rm gc}, Z) = - \frac{G M_{\rm d}}{\sqrt{R_{\rm gc}^2 + \left(a_{\rm d} + \sqrt{Z^2 + b_{\rm d}^2 } \right)^2}} \label{eq:disc}
\end{equation}
where $R_{\rm gc}$ describes the distance from the Galactic center, $Z$ is the vertical distance from the Galactic plane, $G$ is the gravitational constant, $M_{\rm d}$ the mass of the Galactic disk, $a_{\rm d}$ and $b_{\rm d}$ are the scale height parameters of the disk.

The dark matter halo component is typically represented by Navarro-Frenk-White profile \citep[NFW;][]{Navarro96}, given as follows:
\begin{equation}
	\Phi _{\rm halo} (r) = - \frac{G M_{\rm s}}{R_{\rm gc}} \ln \left(1+\frac{R_{\rm gc}}{r_{\rm s}}\right) \label{eq:halo}
\end{equation} 
where $M_{\rm s}$ presents the mass of the dark matter halo of the Milky Way and $r_{\rm s}$ is its radius.

The input parameters needed to perform kinematic analyses and orbit integrations for the two clusters are the central equatorial coordinates ($\alpha, \delta$), mean proper-motion components ($\mu_{\alpha}\cos\delta, \mu_{\delta}$), and distances ($d$). The distances were taken from the isochrones fitting estimates made above in this study. Besides these input parameters, radial velocity data ($V_{\rm R}$) are also required for complete kinematic and orbit analyses. All the input parameters are listed in Table~\ref{tab:Final_table} (on page~\pageref{tab:Final_table}). The mean radial velocities for the two clusters were calculated using the most probable members as selected from the {\it Gaia} DR3 catalog, within the clusters' limiting radii. 13 stars in Rup-1 and 102 for Rup-171 had probabilities $P\geq 0.5$ and were considered in the mean radial velocity calculations. The estimation of mean radial velocities was based on the equations given by \citet{Soubiran18}. These use the weighted average of the data. We determined the mean radial velocities for Rup-1 and Rup-171 as $V_{\rm R}= 10.37 \pm 2.22$ and $V_{\rm R}= 5.32\pm 0.23$  km s$^{-1}$, respectively. These results are within the error of the radial velocity results given by \citet{Soubiran18}, \citet{Dias21} and \citet{Tarricq21}. We adopted the galacto\-cen\-tric distance, circular velocity, and the distance from the Galactic plane of the Sun to be $R_{\rm gc}=8$ kpc, $V_{\rm rot}=220$ km s$^{-1}$ \citep{Bovy15, Bovy12}, and $27\pm 4$ pc \citep{Chen00}, respectively.   

The orbits of Rup-1 and Rup-171 were integrated backward in time with 1 Myr steps up to an age of 3 Gyr from the clusters' present positions in the Galaxy. The output parameters that were estimated from the kinematic and orbit analyses, which are listed in Table~\ref{tab:Final_table}, where $R_{\rm a}$ and $R_{\rm p}$ are apogalactic and perigalactic distances, respectively, and $e$ is eccentricity of the Galactic orbit. $Z_{\rm max}$ is the maximum vertical distance from Galactic plane, ($U$, $V$, $W$) are the space velocity components, and $P_{t}$ is the orbital period. The space velocity components for Rup-1 were derived as $(U, V, W) = (-3.48 \pm 1.44, -10.02 \pm 1.63, -6.21 \pm 0.52)$ km s$^{-1}$, and for Rup-171 as $(-6.39 \pm 0.30, 31.75 \pm 1.43, -45.71 \pm 2.04)$ km s$^{-1}$. \citet{Soubiran18} considered {\it Gaia} DR2 astrometric data \citep{Gaia18} and derived the space velocity components for Rup-1 as $(U, V, W)$ = ($-4.94 \pm 2.63$, $-11.48 \pm 2.53$, $-6.89 \pm 0.63$) and for Rup-171 as $(U, V, W)$ = ($-6.31 \pm 0.22$, $32.38 \pm 0.17$, $-46.36 \pm 0.22$) km s$^{-1}$. These results are based on two cluster members in Rup-1 and 20 in Rup-171. Our findings for the space velocity components are compatible with the results of \citet{Soubiran18}. 

In order to include a correction for the Local Standard of Rest (LSR) we used the space velocity components of \citet{Coskunoglu11}. These are ($U, V, W$) = $(8.83 \pm 0.24$, $14.19 \pm 0.34$, $6.57 \pm 0.21)$ km s$^{-1}$. Using these values we estimated the LSR corrected space velocity components ($(U, V, W)_{\rm LSR}$) as well as total space velocities ($S_{\rm LSR}$) for Rup-1 as $S_{\rm LSR}=6.79\pm 2.28$ and for Rup-171 as $S_{\rm LSR}=60.40\pm 2.55$ km s$^{-1}$ (see also Table~\ref{tab:Final_table}). According to the study of \citet{Bensby10}, it is emphasised that stars with a $S_{\rm LSR}$ of less than 50 km s$^{-1}$ are members of the thin disk, while stars with a $S_{\rm LSR}$ between 70 and 200 km s$^{-1}$ are members of the thick disk population. Accordingly, Rup-1 is a member of the young thin-disk population and Rup-171 is a member of the old thin-disk population.  

We plotted the resulting orbits, as shown in Fig.~\ref{fig:galactic_orbits}. In the figure, the upper and lower left panels represent the side view of the orbits in the $R_{\rm gc} \times Z$ plane for Rup-1 and Rup-171, respectively \citep[see also, ][]{Evcil24, Dursun24, Yucel24} The upper and lower right panels of Fig.~\ref{fig:galactic_orbits} show the distance from the Galactic center as a function of time in the $R_{\rm gc} \times t $ planes for each cluster. The birth and current locations of the two clusters are indicated by yellow-filled triangles and circles, respectively. Pink and green-dashed lines as well as the relevant triangles show the orbits and birth radii of the clusters for upper and lower errors of input parameters. The upper panels of Fig. \ref{fig:galactic_orbits} show that Rup-1 formed outside the solar circle ($R_{\rm Birth}=8.52\pm 0.09$ kpc) and entirely orbits outside the solar circle. The lower panels of Fig. \ref{fig:galactic_orbits} indicate that Rup-171 also formed outside the solar circle ($R_{\rm Birth}=9.95\pm 0.05$ kpc) and enters inside the solar circle during its orbital motion. Another factor affecting the birth position of OCs is the uncertainty in their ages. The uncertainties in the ages of the Rup-1 and Rup-171 OCs investigated in this study were determined as 60 and 200 Myr, respectively. Taking into account the uncertainties in the ages of the two OCs, the change in birth positions for the clusters is shown by the grey regions in the right panels of Fig. \ref{fig:galactic_orbits}. Dynamical orbital analyses show that if uncertainties in cluster ages are considered, birth positions can be varied in the range $8.31\leq R_{\rm gc} \leq 10.42$ kpc for Rup-1 and $6.53 \leq R_{\rm gc}\leq 9.97$ kpc for Rup-171. Considering the uncertainties in cluster ages, it is determined that Rup-1 is likely to form outside the solar circle and Rup-171 is likely to form inside and/or outside the solar circle. 

In this study, the metal abundances of the two OCs and their distances from the Galactic centre at today and at the time of their birth are taken into account. For this purpose, we refer to \citet{Spina22}, who studied the metal abundances of OCs calculated from the analysis of the spectroscopic data from cluster member stars and the distances of these OCs to the Galactic centre. On the ${\rm[Fe/H]} \times R_{\rm gc}$ diagram, it was found that the metal abundances of other OCs located at the same distance from the today positions of Rup-1 and Rup-171 are in the metallicity intervals of $-0.25<{\rm [Fe/H]\,(dex)}<0.13$ and $-0.09<{\rm [Fe/H]\,(dex)}<0.30$, respectively. Considering the metal abundances calculated for Rup-1 ([Fe/H]=-0.09$\pm$0.16 dex) and Rup-171 ([Fe/H]=-0.20$\pm$0.20 dex) in this study, it was determined that Rup-1 is within the metallicity range of \citet{Spina22}, while Rup-171 is outside the expected metallicity range. Nevertheless, this is in much better agreement with the metallicity ranges of \citet{Spina22} $-0.21<{\rm [Fe/H]\,(dex)}<0.18$ and $-0.31 <{\rm [Fe/H]\,(dex)}< 0.09$ when the birth positions of the OCs Rup-1 ($R_{\rm Birth}=8.52\pm0.09$ kpc) and Rup-171 ($R_{\rm Birth}=9.95\pm0.05$ kpc) are considered.

\section{Luminosity and Present-day Mass Functions}
The luminosity function (LF) refers to the distribution of brightness for a group of stars. We took into consideration {\it Gaia} DR3 photometric data in the estimates of LF for each cluster. We selected the main-sequence stars with probabilities $P>0.5$ and located within the limiting radii obtained in Section~\ref{section:rdps}. The number of selected stars and their magnitude range are 72 and $11.3\leq G \leq 20.5$ mag for Rup-1, for Rup-171 the parameters values correspond to 533 stars and $14.25\leq G \leq 20.50$ mag range. However it can be interpreted that due to possible binary star contamination on the cluster's main sequence, it is not likely to detect all binary stars individually in the cluster. Hence, the stars used in luminosity and present-day mass function analyses were considered as single stars. We derived absolute magnitudes $M_{\rm G}$ from apparent $G$ magnitudes by using the equation of $M_{\rm G} = G-5\times \log d +5+A_{\rm G}$, where $d$ is the distance derived in the study and $A_{G}$ is {\it Gaia} photometry based extinction that described by $A_{G}=1.8626\times E(G_{\rm BP}-G_{\rm RP})$ \citep{Cardelli89, Odennell94} (where, $E(G_{\rm BP}-G_{\rm RP})$ is the color excess obtained in the Section~\ref{distance_age}). We plotted the LF distribution for two clusters as shown in Fig.~\ref{fig:luminosity_functions}. The number of stars were calculated for the intervals of 1.0 mag bin. It can be seen from Fig.~\ref{fig:luminosity_functions} that the absolute magnitude ranges lie within the $0\leq M_{\rm G} \leq 9$ mag for Rup-1 (panel a) and $2\leq M_{\rm G} \leq 9$ mag for Rup-171 (panel b). From the Fig.~\ref{fig:luminosity_functions}a it is concluded that Rup-1 retains its massive and low-mass stars because of its young age, whereas Fig.~\ref{fig:luminosity_functions}b shows that most of the massive stars of Rup-171 are evolved due to its old age.            

The present-day mass function (PDMF) provides information about the number density of stars per mass interval, and it is related to the LF. To derive PDMFs, we considered the same stars selected in the LF analyses for each cluster. LFs of Rup-1 and Rup-171 were converted into present-day mass functions (PDMFs) with the aid of {\sc parsec} models \citep{Bressan12} that scaled to the mass fraction ($z$) and age estimated in this study. Using these models, we expressed an absolute magnitude-mass relation with a high degree polynomial equation between $M_{G}$ absolute magnitudes and masses of theoretical main-sequence stars. The derived relation was applied to the observational stars to transform their absolute $M_{\rm G}$ magnitudes into masses. This resulted in the mass range of the main-sequence stars being estimated as $0.75\leq M/ M_{\odot}\leq 2.50$ for Rup-1, and $0.75\leq M/ M_{\odot}\leq 1.50$ for Rup-171. Stellar masses were adjusted to 0.25 mass bins and logarithmic values of the number of stars within each bin were calculated for two clusters. Then we estimated the slope of the mass function by a power law set as by \citet{Salpeter55}: 
\begin{figure}[t!]
	\centering
	\includegraphics[width=1\linewidth]{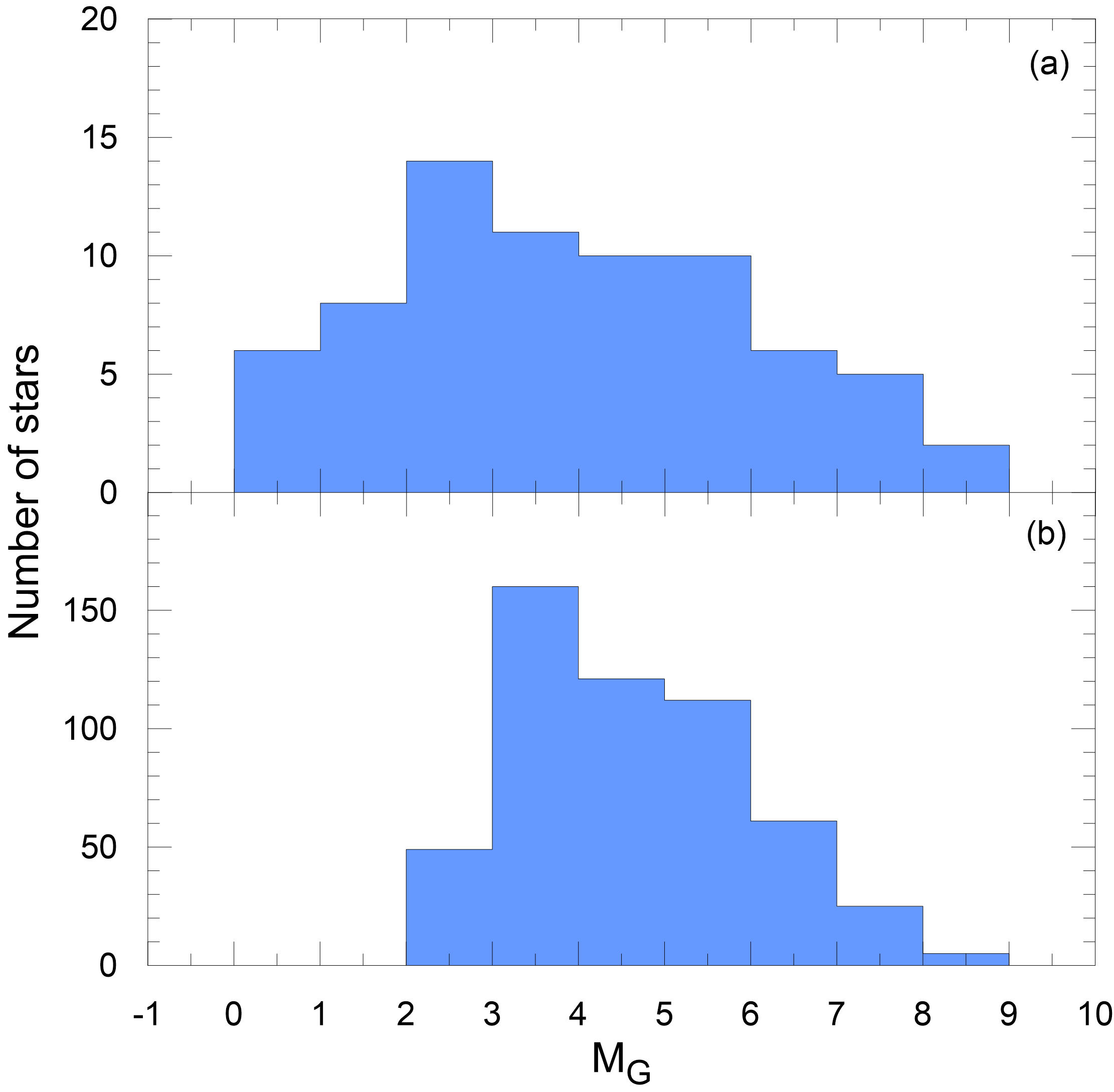}
	\caption{
		The LF histograms for Rup-1 (a) and Rup-171 (b). The distributions present the main-sequence stars with probabilities $P\geq 0.5$ and those located in the limiting radii of the clusters in each magnitude bin.}
  \label{fig:luminosity_functions}
\end{figure}
\begin{figure}[h!]
	\centering
	\includegraphics[width=0.98\linewidth]{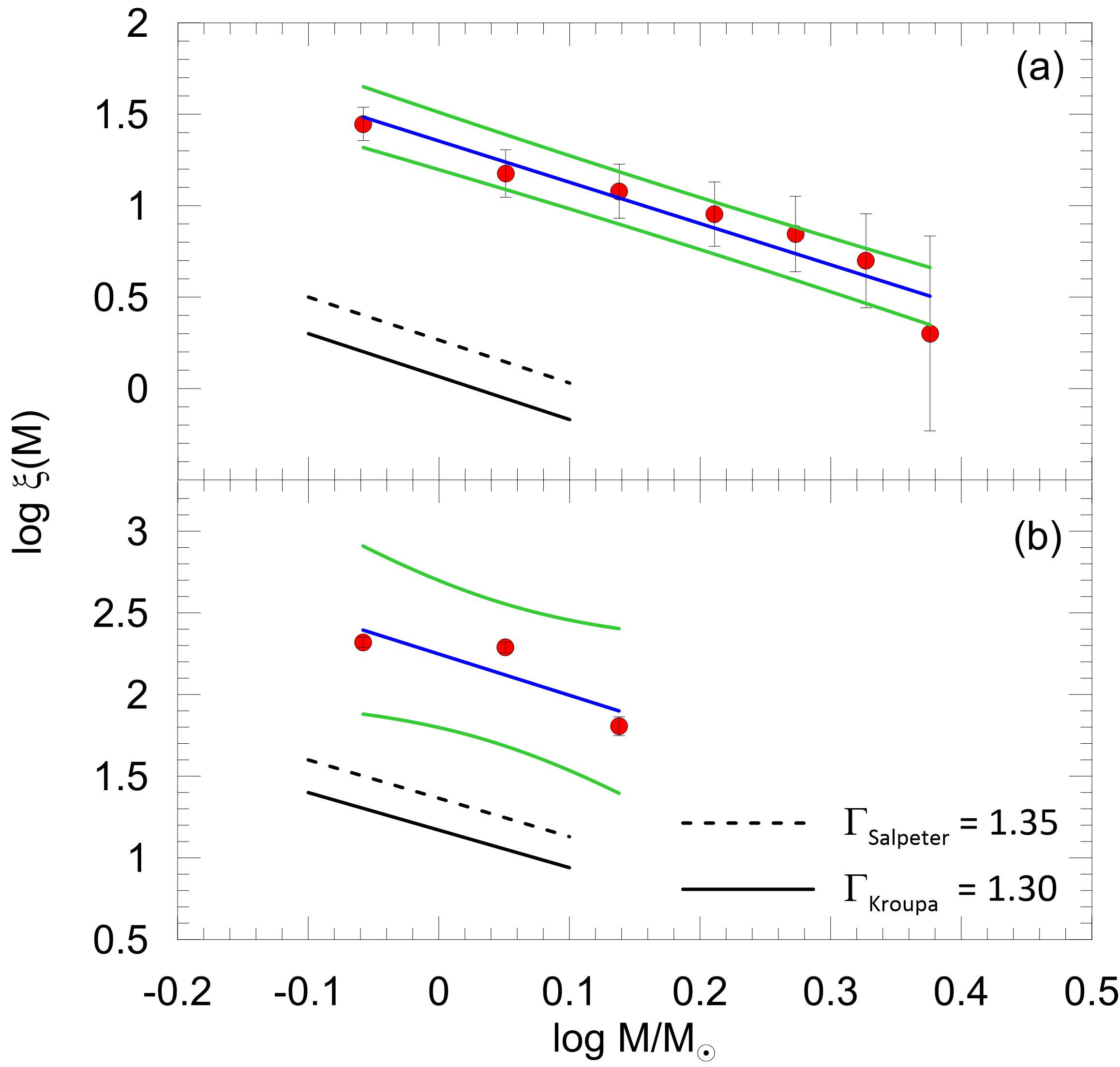}
	\caption{
		PDMF fits of Rup-1 (a) and Rup-171 (b) derived from the stars with probabilities $P\geq 0.5$ and those located in the limiting radii of the clusters (red circles). The blue and green lines show the PDMFs and their errors, which represent the $\pm1\sigma$ prediction levels. Dashed and solid lines identify the mass function of \citet{Salpeter55} and \citet{Kroupa01, Kroupa02}, respectively.}
  \label{fig:mass_functions}
\end{figure}

\begin{equation}
	{\log(\frac{dN}{dM})=-(1+\Gamma)\times \log(M)+{\rm constant}}
\end{equation} 
here $dN$ symbolizes the number of stars in a mass bin $dM$, $M$ represents the central mass of the relevant bin and $\Gamma$ is the slope of the function. The best-fit PDMFs are plotted in Fig.~\ref{fig:mass_functions}. The derived PDMFs are $\Gamma = 1.26 \pm 0.32$ for Rup-1 and $\Gamma = 1.53 \pm 1.49$ for Rup-171, which agree with the value of $\Gamma=1.35$ given by \citet{Salpeter55} and the value of $\Gamma=1.30$ provided by \citet{Kroupa01, Kroupa02} within error. In addition to this, the total mass of the clusters ($M_{\rm tot}$) and mean masses of the member stars ($\langle m \rangle$) for Rup-1 and Rup-171 were calculated as 99 and 1.33 $M/M_{\odot}$, and as 623 and 1.05 $M/M_{\odot}$, respectively. Moreover, it is found that the uncertainties in the metal abundances of the two OCs can lead to a change of at most 0.05 $M/M_{\odot}$ in the stellar mass calculations. This has a no direct impact on the determination of the mass functions of two OCs analysed.


\section{Summary and Conclusion}
We present a comprehensive study of the two OCs Ruprecht 1 and Ruprecht 171 taking into account CCD {\it UBV} photometric as well as {\it Gaia} DR3 astrometric, photometric, and spectroscopic data. Analyses of fundamental astrophysical parameters were performed by using the {\it UBV} data, whereas the estimation of distances and ages, orbit integrations, and structural analyses were based on the {\it Gaia} DR3 data. The main results are listed in Table~\ref{tab:Final_table} and summarized as follows:

\begin{enumerate}
	\item{RDP analyses utilized the {\it Gaia} DR3 data gathered in 25 arcmin radii areas about the cluster centers. We fitted King profiles to the stellar densities, obtaining through visual inspection the limiting radius of Rup-1 as $r_{\rm lim}=7'$ and for Rup-171 as $r_{\rm lim}=10'$. These values correspond to the limiting radii for Rup-1 and Rup-171 being 2.99 pc and 4.39 pc, respectively.}
	
	\item{The membership probability calculation was based on {\it Gaia} DR3 proper motion components, trigonometric parallaxes, and their uncertainties. We adopted as possible cluster members the stars with membership probabilities $P\geq 0.5$. To perform {\it UBV} data-based analyses, the membership probability values of the same stars in the {\it Gaia} DR3 and {\it UBV} catalog were cross-matched. We made a selection of the most probable member stars for these two catalogs separately:
		
		\begin{enumerate}[a)]
			\item{For {\it UBV} data, we considered binary star contamination on main-sequence stars that lie within the clusters' limiting radii. We fitted intrinsic ZAMS to $V\times (B-V)$ CMDs of the two clusters and shifted it $\Delta V=0.75$ mag towards the brighter stars. In addition to this criteria, for {\it UBV} data, we selected the stars with membership probabilities $P\geq 0.5$ and brighter than faint $V$ magnitude limit and identified 36 and 115 most probable member stars for Rup-1 and Rup-171, respectively.}
			
			\item{For {\it Gaia} DR3 data, we selected the stars located within the clusters' limiting radii and those brighter than faint $G$ magnitude limit and with membership probabilities $P\geq 0.5$ as the most probable members. Hence, for {\it Gaia} DR3 data, we estimated the number of most probable member stars to be 74 and 596 for Rup-1 and Rup-171, respectively. Consequently, {\it UBV} and {\it Gaia} data-based analyses were performed considering the member stars identified from the relevant catalog.} 
   
		\end{enumerate}
	The number of most probable cluster stars are different between the {\it UBV} and {\it Gaia} samples. The limited field of view of the {\it UBV} photometric observations and/or exposure times may influence the number of detected stars. To avoid `loss' of stars that may be caused by these reasons and to improve parameter determinations such as for age, LF, and PDMF we therefore considered also {\it Gaia} data for its larger field of view.}

		\item{Mean proper-motion components for Rup-1 were calculated as ($\mu_{\alpha}\cos \delta, \mu_{\delta}) = (-0.287 \pm 0.003, -0.903 \pm 0.003$) mas yr$^{-1}$ and for Rup-171 as ($\mu_{\alpha}\cos \delta, \mu_{\delta}) = (7.720 \pm 0.002, 1.082 \pm 0.002$) mas yr$^{-1}$.}  
		
		\item{The mean trigonometric parallax was derived for Rup-1 as $\varpi_{\rm Gaia}= 0.649 \pm 0.027$ mas, and for Rup-171 as $\varpi_{\rm Gaia}= 0.631 \pm 0.042$ mas. Using the linear equation of $\varpi \: ({\rm mas})=1000/d \: ({\rm pc})$, we calculated trigonometric parallax-based distances ($d_{\rm\varpi}$) for Rup-1 and Rup-171 as $1541 \pm 64$ pc and $1585 \pm 106$ pc, respectively.}
		
		\item{We identified the four most probable BSSs in Rup-171 within the 5 arcmin area from the cluster's center. Three of these stars were previously identified in the study of \citet{Jadhav21}.}  
		
		\item{The color excesses and photometric metallicities of the two clusters were derived separately from $(U-B)\times (B-V)$ TCDs. The $E(B-V)$ color excess and [Fe/H] photometric metallicity are $0.166 \pm 0.022$ mag and $-0.09 \pm 0.16$ dex for Rup-1, respectively. These values correspond to $0.301 \pm 0.027$ mag and $-0.20 \pm 0.20$ dex for Rup-171.}	
		
		\item{The distance and age of the two clusters were estimated simultaneously on {\it UBV} and {\it Gaia} DR3 data-based CMDs. Keeping as constants the derived color excesses and metallicities, we estimated apparent distance moduli, distance, and age of Rup-1 as $\mu_{\rm V}=11.346\pm 0.083$ mag, $d=1469\pm 57$ pc, and $t=580\pm 60$ Myr, respectively. Similarly $\mu_{\rm V}=11.819\pm 0.098$ mag, $d=1509\pm69$ pc, and $t=2700\pm 200$ Myr were obtained for Rup-171. According to the {\it Gaia} DR3 data-based results, the best solution of $E(G_{\rm BP}-G_{\rm RP})$ was achieved when we consider the equation of $E(G_{\rm BP}-G_{\rm RP})= 1.29\times E(B-V)$ of \citet{Wang19}.}
		
		\item{The results of space velocities and Galactic orbital parameters indicated that Rup-1 belongs to the young thin-disk population, whereas Rup-171 is a member of the old thin-disk population. Also, we concluded that Rup-1 and Rup-171 formed outside the solar circle with the birth radii of $8.52\pm 0.09$ pc and $9.95\pm 0.05$ kpc, respectively, but only Rup-1 entirely orbits outside the solar circle. Considering the uncertainties in cluster ages, it is determined that Rup-1 is likely to form outside the solar circle and Rup-171 is likely to form inside and/or outside the solar circle.}
		
		\item{Results of PDMFs were found as $\Gamma=1.26\pm 0.32$ and $\Gamma=1.53 \pm 1.49$ for Rup-1 and Rup-171, respectively, which are in good agreement with the value of \citet{Salpeter55}. Also, the total masses of the clusters and mean masses of the member stars for Rup-1 and Rup-171 were calculated as 99 and 1.33 $M/M_{\odot}$, and as 623 and 1.05 $M/M_{\odot}$, respectively.}
\end{enumerate}

The study of the OCs analysed in this paper with {\it Gaia} DR3 data and different filter sets minimised the degeneracy between the parameters by allowing the basic astrophysical parameters to be calculated with independent methods. This will contribute to the study understanding of the Galactic structure and the understanding of the chemo-dynamic evolution of the Galactic disk, as a result of investigating a large number of OCs with the same method.

\begin{table}
	\setlength{\tabcolsep}{3pt}
	\renewcommand{\arraystretch}{1.5}
	\fontsize{7pt}{7pt}\selectfont
	\centering
	\caption{Fundamental parameters of Rup-1 and Rup-171.}
	\begin{tabular}{@{}lrr@{}}
		\hline \\[-3ex]
		Parameter & Rup-1 & Rup-171 \\
		\hline
		($\alpha,~\delta)_{\rm J2000}$ (sexagesimal)& 06:36:20.20, $-$14:09:25.25 & 18:32:02.90, $-16$:03:43.00 \\
		($l, b)_{\rm J2000}$ (decimal)              & 223.9600, $-09.6918$         & 016.4520, $-03$.0891       \\
		$f_{0}$ (stars arcmin$^{-2}$)               & $51.550\pm 3.132$            & $7.610 \pm 0.973$          \\
		$r_{\rm c}$ (arcmin)                        & $0.254 \pm 0.016$            & $3.297 \pm 0.920$          \\
		$f_{\rm bg}$ (stars arcmin$^{-2}$)          & $7.573 \pm 0.136$            & $148.411 \pm 2.487$        \\
		$r_{\rm lim}$ (arcmin)                      & 7                            & 10                         \\
		$r$ (pc)                                    & 2.99                         & 4.39                       \\
		$\mu_{\alpha}\cos \delta$ (mas yr$^{-1}$)   & $-0.287 \pm 0.003$           & $7.720 \pm 0.002$          \\
		$\mu_{\delta}$ (mas yr$^{-1}$)              & $-0.903 \pm 0.003$           & $1.082 \pm 0.002$          \\
		Cluster members ($P\geq0.5$)                & 74                           & 596                        \\
		$\varpi$ (mas)                              & $0.649 \pm 0.027$            & $0.631 \pm 0.042$          \\
		$E(B-V)$ (mag)                              & $0.166 \pm 0.022$            & $0.301 \pm 0.027$          \\
		$E(U-B)$ (mag)                              & $0.121 \pm 0.016$            & $0.221 \pm 0.019$          \\
		$A_{\rm V}$ (mag)                           & $0.515 \pm 0.068$            & $0.933 \pm 0.083$          \\
		$[{\rm Fe/H}]$ (dex)                        & $-0.09 \pm 0.16$             & $-0.20 \pm 0.20$           \\
  	    $z$                                         & $0.012 \pm 0.003$            & $0.010 \pm 0.004$          \\
		Age (Myr)                                   & $580 \pm 60$                 & $2700 \pm 200$             \\
		$V-M_{\rm V}$ (mag)                         & $11.346 \pm 0.083$           & $11.819 \pm 0.098$         \\
		$d_{\rm iso}$ (pc)                          & $1469 \pm 57$                & $1509 \pm 69$              \\
		$(X, Y, Z)_{\odot}$ (pc)                    & ($-1042$, $-1005$, $-247$)   & ($1445$, 427, $-81$)       \\
		$R_{\rm gc}$ (kpc)                          & 9.10                         & 6.57                       \\
		PDMF slope                                  & $1.26\pm 0.32$               & $1.53 \pm 1.49$            \\
		$M_{\rm tot}$ ($M/M_{\odot}$)               & 99                           & 623                        \\ 
		$V_{\rm R}$ (km s$^{-1}$)                   & $10.37 \pm 2.22$             & $5.32 \pm 0.23$            \\
		$U_{\rm LSR}$ (km s$^{-1}$)                 & $5.35 \pm 1.46$              & $2.44 \pm 0.38$            \\
		$V_{\rm LSR}$ (km s$^{-1}$)                 & $4.17 \pm 1.66$              & $45.94 \pm 1.47$           \\
		$W_{\rm LSR}$ (km s$^{-1}$)                 & $0.36 \pm 0.56$              & $-39.14 \pm 2.05$          \\
		$S_{_{\rm LSR}}$ (km s$^{-1}$)              & $6.79 \pm 2.28$              & $60.4 \pm 2.55$            \\
		$R_{\rm a}$ (kpc)                           & $10.42\pm 0.07$              & $9.97 \pm 0.05$            \\
		$R_{\rm p}$ (kpc)                           & $8.31 \pm 0.01$              & $6.54 \pm 0.07$            \\
		$z_{\rm max}$ (pc)                          & $248 \pm 8$                  & $661 \pm 44$               \\
		$e$                                         & $0.113 \pm 0.004$            & $0.208 \pm 0.008$          \\
		$P_{t}$ (Myr)                               & $266 \pm 1$                  & $234 \pm 1$                \\
		$R_{\rm Birth}$ (kpc)                       & $8.52 \pm 0.09$              & $ 9.95\pm 0.05$            \\
		\hline
	\end{tabular}%
	\label{tab:Final_table}%
\end{table}%

\section*{Acknowledgments}
This study has been supported in part by the \fundingAgency{Scientific and Technological Research Council (TÜBİTAK)} \fundingNumber{122F109}. The observations of this publication were made at the National Astronomical Observatory, San Pedro M{\'a}rtir, Baja California, M{\'e}xico, and the authors thank the staff of the Observatory for their assistance during these observations. The authors express their sincere gratitude to the anonymous referee for providing invaluable feedback and suggestions that have significantly enhanced the readability and overall quality of the paper. This research has made use of the WEBDA database, operated at the Department of Theoretical Physics and Astrophysics of the Masaryk University, and also made use of NASA's Astrophysics Data System.  The VizieR and Simbad databases at CDS, Strasbourg, France were invaluable for the project as were data from the European Space Agency (ESA) mission {\it Gaia}\footnote{https://www.cosmos.esa.int/gaia}, processed by the {\it Gaia} Data Processing and Analysis Consortium (DPAC)\footnote{https://www.cosmos.esa.int/web/gaia/dpac/consortium}. Funding for DPAC has been provided by national institutions, in particular the institutions participating in the {\it Gaia} Multilateral Agreement. IRAF was distributed by the National Optical Astronomy Observatory, which was operated by the Association of Universities for Research in Astronomy (AURA) under a cooperative agreement with the National Science Foundation. PyRAF is a product of the Space Telescope Science Institute, which is operated by AURA for NASA.

\subsection*{Author contributions}

\textbf{Conception/Design of study}: Hikmet \c{C}akmak, Sel\c {c}uk Bilir, Talar Yontan, Timothy Banks;\\ 
\textbf{Data Acquisition}: Hikmet \c{C}akmak, Seliz Ko\c{c}, Hülya Er\c{c}ay, Talar Yontan, Ra\'ul Michel; \\
\textbf{Data Analysis/Interpretation}: Hikmet \c{C}akmak, Sel\c{c}uk Bilir, Talar Yontan, Timothy Banks, Ra\'ul Michel, Seliz Ko\c{c}, Hülya Er\c{c}ay;\\
\textbf{Drafting Manuscript}: Sel\c{c}uk Bilir, Talar Yontan, Hikmet \c{C}akmak, Timothy Banks, Ra\'ul Michel, Esin Soydugan;\\
\textbf{Critical Revision of Manuscript}: Hikmet \c{C}akmak, Sel\c{c}uk Bilir, Talar Yontan, Timothy Banks, Ra\'ul Michel, Esin Soydugan; \\
\textbf{Final Approval and Accountability}: Sel\c{c}uk Bilir, Talar Yontan, Hikmet \c{C}akmak.
\subsection*{Financial disclosure}

None reported.

\subsection*{Conflict of interest}

The authors declare no potential conflict of interests.

\bibliography{refs_cakmak}%

%
\end{document}